\documentclass[10pt, conference, compsocconf]{IEEEtran}

\usepackage{hyperref}

\usepackage{amsmath,amssymb,amsfonts}
\usepackage{algorithmic}
\usepackage{graphicx}
\usepackage{tabularx}
\usepackage{textcomp}
\usepackage{xcolor}
\usepackage{booktabs}
\usepackage{cleveref}
\usepackage[vlined,ruled,linesnumbered]{algorithm2e}
\usepackage{mathtools}
\usepackage{subcaption}
\usepackage[normalem]{ulem}

\newcommand{\tool}[0]{SentiNet}

\crefname{algocf}{alg.}{algs.}
\Crefname{algocf}{Algorithm}{Algorithms}

\SetKwInOut{Input}{in}
\SetKwInOut{Output}{out}
\SetKwFunction{SelectiveSearch}{SelectiveSearch}
\SetKwFunction{InertPattern}{InertPattern}
\SetKwFunction{Overlay}{Overlay}
\SetKwFunction{MaskGradCAM}{MaskGradCAM}
\SetKwFunction{MaskGeneration}{MaskGeneration}
\SetKwFunction{ClassProposal}{ClassProposal}
\SetKwFunction{OutPts}{OutPts}
\SetKwFunction{ApproximateCurve}{ApproximateCurve}
\SetKwFunction{COBYLA}{COBYLA}
\SetKwFunction{TopConf}{TopConf}

\begin{document}

\title{\tool: Detecting Localized Universal Attacks Against Deep Learning Systems}

 \author{\IEEEauthorblockN{Edward Chou}
	\IEEEauthorblockA{Stanford University}
	\and
	\IEEEauthorblockN{Florian Tram\`er}
	\IEEEauthorblockA{Stanford University}
	\and
	\IEEEauthorblockN{Giancarlo Pellegrino}
	\IEEEauthorblockA{CISPA Helmholtz Center for Information Security}}

\maketitle

\begin{abstract}
\tool{} is a novel detection framework for \emph{localized universal attacks} on neural networks. These attacks restrict adversarial noise to contiguous portions of an image and are reusable with different images---constraints that prove useful for generating physically-realizable attacks.  Unlike most other works on adversarial detection, \tool{} does not require training a model or preknowledge of an attack prior to detection.  Our approach is appealing due to the large number of possible mechanisms and attack-vectors that an attack-specific defense would have to consider.  By leveraging the neural network's susceptibility to attacks and by using techniques from model interpretability and object detection as detection mechanisms, \tool{} turns a weakness of a model into a strength.  We demonstrate the effectiveness of \tool{} on three different attacks---i.e., data poisoning attacks, trojaned networks, and adversarial patches (including physically realizable attacks)---
and show that our defense is able to achieve very competitive performance metrics for all three threats. Finally, we show that \tool{} is robust against strong adaptive adversaries, who build adversarial patches that specifically target the components of \tool{}'s architecture.
\end{abstract}

\section{Introduction}



Deep neural networks are susceptible to a variety of attacks aimed at causing misclassifications~\cite{szegedy2013boxlbfgs,goodfellow2014fgsm, brown2017adversarialpatch}.
As deep learning models are also often re-used (e.g., via transfer learning~\cite{bengio2012deep}), model vulnerabilities---whether inherent or purposefully inserted by a malicious party~\cite{Liu2018TrojaningAO}---can easily affect a large number of systems.
This has severe implications for the trustworthiness of deep learning models in security-critical decision-making situations.
Defending against these attacks is challenging due to the wide variety of possible attack mechanisms and vectors, especially for models operating in the visual domain. In this work, we explore \emph{localized} and \emph{universal} attacks on visual classifiers and introduce \tool{}, a robust defense which detects adversarial inputs without requiring any specific model re-training or prior knowledge of the attack vector.

We focus on attacks that are \emph{localized}, i.e., the adversarial region is constrained to a small contiguous portion of an image, and \emph{universal}, i.e., attacks are image-agnostic. 
Inputs with these properties have proven useful for instantiating robust and physically realizable attacks that take the form of an adversarial object or sticker placed inside a visual scene~\cite{brown2017adversarialpatch, Liu2018TrojaningAO,evtimov2017robust,eykholt2018physical}. These classes of attacks typically use \emph{unbounded} perturbations (i.e., without any specific $\ell_\infty$ or $\ell_2$ constraint as in most digital attacks~\cite{szegedy2013boxlbfgs, goodfellow2014fgsm}), to ensure that the attacks are robust to changes in viewpoint, lighting and other physical artifacts. 
Several such physical attacks have been demonstrated, with the goal of causing misclassifications when applied to arbitrary images with different class labels~\cite{brown2017adversarialpatch, Liu2018TrojaningAO}.
A drawback of localized attacks is that they are generally visible and detectable by the human eye, but there are many situations where attacks can be deployed in autonomous settings or carefully disguised~\cite{brown2017adversarialpatch}.

A prospective defender must first consider that the model being protected may have been compromised prior to being deployed.
An attack can originate from the source of the network provider, as in the case with data poisoning attacks~\cite{gu2017badnets}, or can be intercepted and modified, as with trojaning attacks~\cite{Liu2018TrojaningAO}.
Even if a network is integrity protected, adversaries can still generate localized and universal attacks that will affect the model at test time, via adversarial examples~\cite{brown2017adversarialpatch}.
Altogether, this creates an extremely difficult security setting where vulnerabilities are easily distributed, where adversaries target properties inherent to neural network systems that cannot be removed, and where attacks might be too diverse in appearance for a signature-based scheme.

Our goal is to create a defense against localized universal attacks that is attack-vector agnostic.
To this end, we identify unifying and necessary features of successful localized universal attacks---including physically realizable attacks---and develop \tool{}, a novel technique that exploits these attack behaviors to detect them.
We start from the observation that localized universal attacks are designed to be robust to a variety of artifacts while generalizing to a large distribution of inputs (e.g., the adversarial patch of~\cite{brown2017adversarialpatch} is designed to work when applied to \emph{any} input image). 
Our first insight is that a localized universal attacks' success relies on the use of ``salient'' features that strongly affect the model's classification on many different inputs. We thus consider techniques from \emph{model interpretability} and \emph{object detection} to discover highly salient contiguous regions of an input image. As we show, these techniques uncover adversarial image regions, as well as benign ones that strongly affect classification. In a second step, we exploit the strong robustness and generalization properties of malicious regions to distinguish them from benign regions with high saliency. Specifically, we overlay extracted regions on a large number of held-out clean images and test how often this results in a misclassification. Malicious regions are much more likely than benign regions to generate misclassifications and are thus detected by \tool{}. As we show in our evaluation of \tool{}, mounting an attack that evades detection requires lowering an adversarial region's saliency to a point where the region is no longer universal (i.e., it fails with high probability on random inputs)---even for a adaptive adversary with knowledge of \tool{}'s architecture. We also validate \tool{}'s effectiveness in a realistic physical setting, where it successfully detects a printed adversarial patch~\cite{brown2017adversarialpatch} with high reliability.

\paragraph{Contributions} To summarize, this paper makes the following contributions:


\begin{itemize}
	\item We propose \tool{}, a new architecture that protects a neural network by using the same model to detect localized universal attacks.

	\item To the best of our knowledge, \tool{} is the first architecture that does not rely on prior knowledge of the specific attack vector used to carry the attack, e.g., trojaned networks, poisoned networks, and adversarial patches---including physically realizable attacks.

	\item \tool{} uses a novel approach to detect a potential attack region using techniques developed for model visualization and object detection and feeds the attack deployed on multiple test images back to the network to perform attack classification.
	
	\item We evaluate \tool{} to protect pre-existing compromised and uncompromised networks against three known attack vectors, i.e., backdoors for poisoned networks, triggers for trojaned networks, and adversarial patches for uncompromised networks. We show that \tool{} can protect neural networks successfully, with an average true positive rate of 96.22\% and an average true negative rate of 95.36\%.

	\item We further evaluate \tool{} against a fully adaptive white-box adversary and present seven attacks against \tool{}'s core components. We show that \tool{} is resistant against strong adversaries by demonstrating the robustness of each individual component.


\end{itemize}

\paragraph{Paper Organization} In \Cref{sec:background}, we present the architecture of deep learning systems and \tool{}'s threat model. Then, in \Cref{sec:sentinet}, we present \tool{} and its architecture. In \Cref{sec:evaluation}, we evaluate \tool{} against localized universal attacks taken from the literature, as well as against physically printed adversarial patches. In \Cref{sec:adaptive_attacks}, we assess \tool{} against adaptive attackers trying to fool \tool{}'s internal components. Next, in \Cref{sec:discussion}, we discuss our results. Finally, in \Cref{sec:related_work}, we cover related works, and in \Cref{sec:conclusion}, we provide our conclusions.


\section{Background}
\label{sec:background}



\subsection{Deep Learning Systems}
\label{sec:dl_systems}


\emph{Deep learning} (DL) is a branch of machine learning which focuses on multi-layered artificial neural networks~\cite{lecun2015deep}. A neural network can be defined as a standard machine learning function $f_m$, which given an input $x$ returns a prediction $y$ and prediction-confidence $conf$; i.e. $(y, conf)=f_m(x)$.  Convolutional Neural Networks (CNN)~\cite{lecun1995convolutional} are a specific architecture of neural network primarily targeted at computer vision tasks. Common CNN architectures include VGG-16~\cite{simonyan2018vgg16}, ResNets~\cite{he2015resnet} and Inception models~\cite{szegedy2014inception}.


\emph{DL systems} are computer systems that rely on neural networks to perform specific tasks. At its essence, a DL system takes in an input, computes an output value, and makes a prediction based on the output. The analysis performed by a DL system is often a classification task such as a face-based user authentication mechanisms, voice recognition, or voice-to-text system. Recently, DL systems have been proposed for autonomous vehicles and digital assistants~\cite{zhang2018dlreview}.




\subsection{Threat Model}
\label{sec:threat_model}

In this work, we assume a scenario where a DL system uses a pre-trained deep CNN model to classify sensor data. Within this scenario, the goal of the adversary is to hijack the prediction of the model by providing a malicious image to the CNN model. In \Cref{sec:lua}, we present the properties of the malicious inputs considered in this paper, and in \Cref{sec:attack_vectors}, we present existing techniques to mount the attacks. Additionally, we assume an adversary that is aware of the defense mechanisms, and that can generate malicious images to bypass \tool{}. In \Cref{sec:lua}, we present such an adversary.

\begin{figure*}
	\centering
    \begin{subfigure}[b]{0.32\linewidth}
        \centering
        \includegraphics[width=0.45\columnwidth]{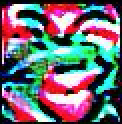}
        \includegraphics[width=0.45\columnwidth]{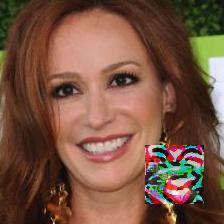}
        \caption{Trojan trigger.}
        \label{fig:trigger}
    \end{subfigure}\hfill
    \vspace{0.2cm}
    \begin{subfigure}[b]{0.32\linewidth}
        \centering
        \includegraphics[width=0.45\columnwidth]{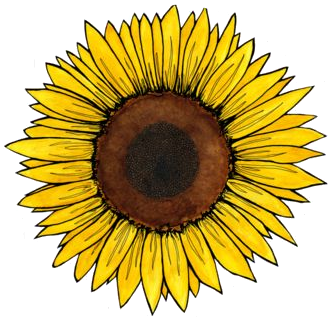}
        \includegraphics[width=0.45\columnwidth]{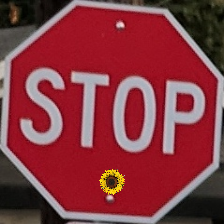}
        \caption{Backdoor.}
        \label{fig:backdoor}
    \end{subfigure}\hfill
    \vspace{0.2cm}
    \begin{subfigure}[b]{0.32\linewidth}
        \centering
        \includegraphics[width=0.45\columnwidth]{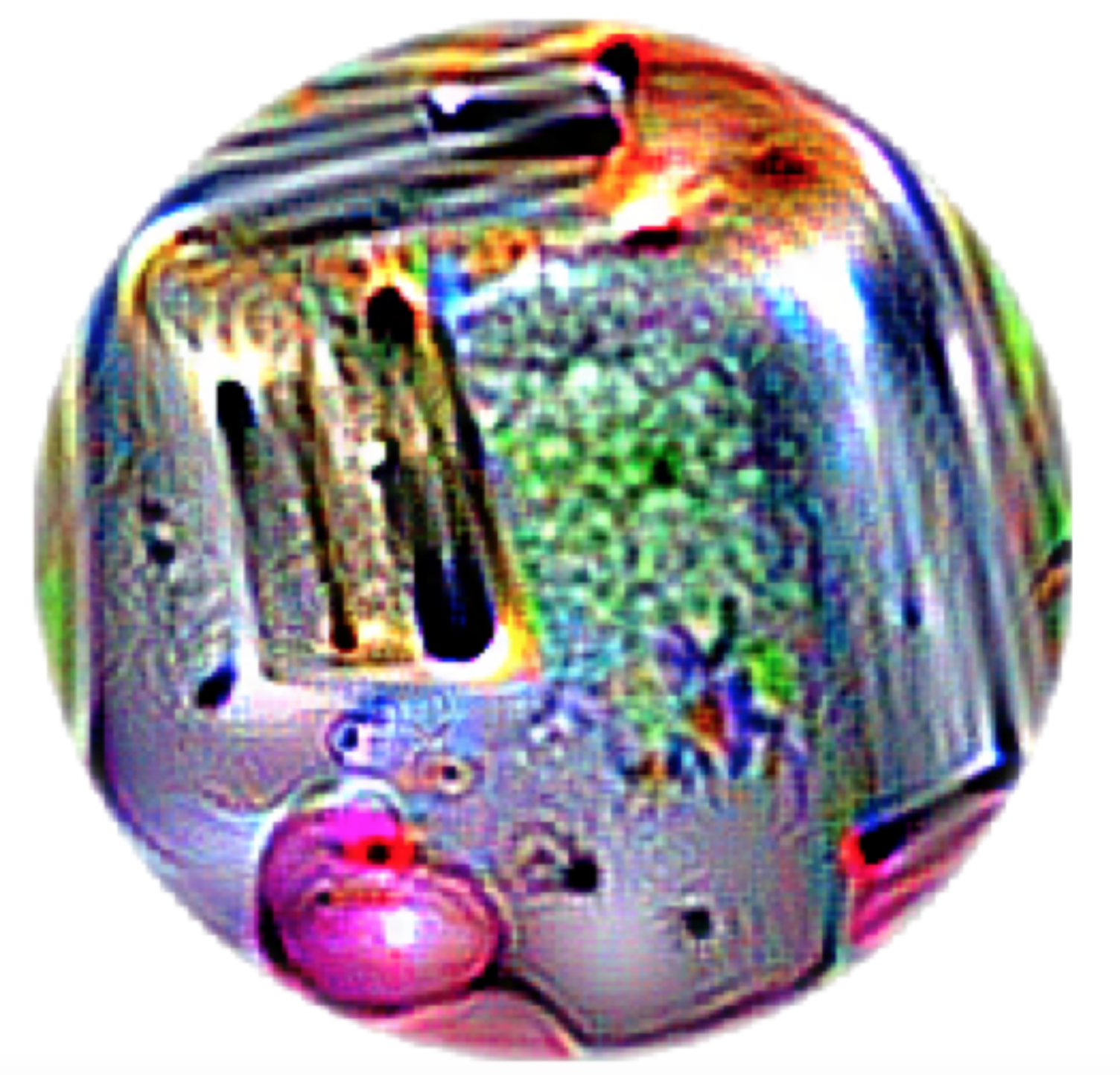}
        \includegraphics[width=0.45\columnwidth]{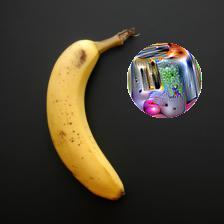}
        \caption{Adversarial patch.}
        \label{fig:adv_patch}
	\end{subfigure}
    \caption{Examples of localized universal attacks. (a) Trojan trigger for trojaned VGG-16 model for face recognition~\cite{Liu2018TrojaningAO}. (b) Backdoor for poisoned Faster-RCCN for road signs detection~\cite{gu2017badnets}. (c) Adversarial patch for uncompromised VGG-16 for ImageNet~\cite{brown2017adversarialpatch}.}
    \label{fig:lua}
\end{figure*}

\subsubsection{Localized Universal Attacks}
\label{sec:lua}

The goal of the adversary is to hijack the prediction of the model to gain control of the actions performed by the DL system. To achieve this goal, the adversary can craft a malicious object and place it in the input image, e.g., inside the visual scene of the sensors. The malicious objects considered in this paper are \emph{localized}, i.e., the object is constrained to a small contiguous region of the image. The objects of this paper are also \emph{universal}, i.e., the object is created once and reused many times with different input images. \Cref{fig:lua} shows three examples of universal, localized malicious objects. 

\subsubsection{Attack Vectors}
\label{sec:attack_vectors}

The adversary can construct and use malicious objects in different ways, that we review in this section. For example, the adversary can ``alter'' the behavior of the network to respond to a malicious object before its deployment in a DL system. Examples of these attacks are trojaning~\cite{ji2018model, Liu2018TrojaningAO} or poisoning attacks~\cite{gu2017badnets}. Alternatively, the adversary can also cause misclassifications to uncompromised models by crafting ad-hoc malicious object, e.g., adversarial patches~\cite{sharif2016glasses,kurakin2016adversarial, brown2017adversarialpatch,evtimov2017robust,eykholt2018physical,athalye2017synthesizing}.


\paragraph{Compromised Models} One way to mount localized universal attacks is to compromise the network before deployment, by establishing a strong response between a class and a malicious object. Over the past few years, two attack vectors have emerged, i.e., data poisoning attacks and trojaning attacks. In a data poisoning attack, the adversary compromises the network at training time by inserting malicious inputs, i.e., benign images with malicious objects called \emph{backdoors} (see \Cref{fig:backdoor}), in the training set~\cite{gu2017badnets}. In a model trojaning attack, the adversary compromises the network by modifying the weights of selected neurons to respond to a specific \emph{trojan trigger}~\cite{ji2018model, Liu2018TrojaningAO} (see \Cref{fig:trigger}).


\paragraph{Vulnerabilities of Uncompromised Models} Even if the networks are integrity protected, an adversary can hijack models' predictions by exploiting the inherent instability of neural networks when processing maliciously-crafted inputs. Recent works have shown that adversaries can create localized universal malicious objects, e.g., printed patches~\cite{sharif2016glasses,brown2017adversarialpatch,eykholt2018physical,evtimov2017robust} or 3D objects~\cite{athalye2017synthesizing} (see, e.g., \Cref{fig:adv_patch}), that can fool a model under real-world conditions such as lighting, sensor noise, and rotation.


\subsubsection{Adaptive Adversary} Over the past years, many ideas have been proposed to protect neural networks from attacks. While some new ideas have moderately increased the robustness of neural networks (e.g., strong adversarial training~\cite{madry2018towards}), many do not adequately protect networks against adversaries aware of the specific defense mechanism being used (see, e.g.,~\cite{carlini2017adversarial,athalye2018obfuscated,tramer2020adaptive}). Accordingly, in this paper, we assume a strong white-box adversary that is fully aware of \tool{}, its architecture and mechanisms. 

\begin{figure*}[!htbp]
    \centering
    \includegraphics[width=1\textwidth]{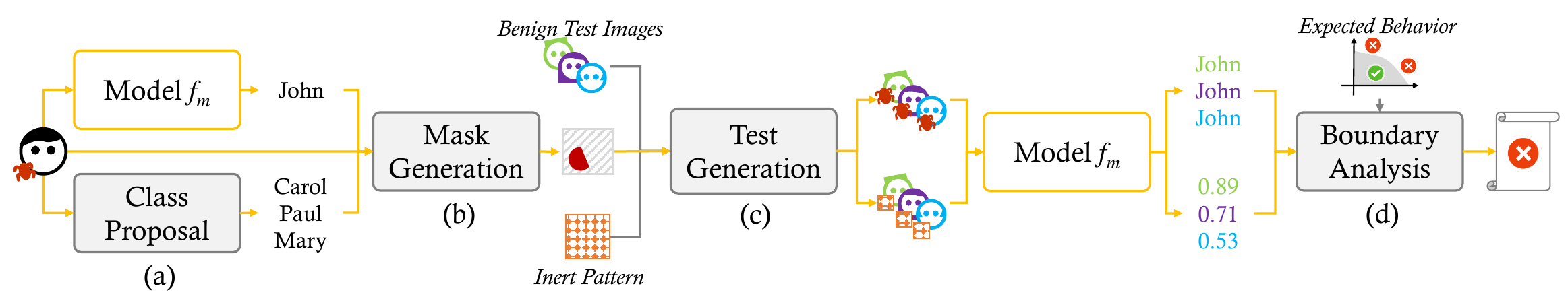}
    \caption{Overview of the \tool{} architecture to protect a model $f_m$.  The output and class proposals of an input are used to generate masks and image tests, which are then fed back into $f_m$ to generate values for boundary analysis and attack classification.}
    \label{fig:approach_overview}
\end{figure*}

\section{\tool}
\label{sec:sentinet}

In this section, we present \tool{}. The goal of \tool{} is to identify adversarial inputs that will hijack the prediction of the neural network without assuming the knowledge of the attack vector beforehand, e.g., exploiting vulnerabilities of compromised and uncompromised networks. The core insight of \tool{} is to use the very behavior of adversarial misclassification to detect an attack. 

The architecture of \tool{} is shown in \Cref{fig:approach_overview}. First, \tool{} uses techniques from model interpretability and object detection to extract from an input scene $x$ those regions that most highly influence the model prediction $y$ (\Cref{sec:attack_localization}). These regions likely contain the malicious object (if present) as well as benign salient regions. Then, \tool{} applies these extracted regions on a set of benign test inputs and observes the behavior of the model. Finally, \tool{} compares the synthetic behaviors with the known behavior of the model on benign inputs, to detect prediction hijacking (\Cref{sec:fuzzing}).

\subsection{Adversarial Object Localization}
\label{sec:attack_localization}

The first phase of our approach intends to localize on the given input the regions that might contain malicious objects. The idea is to identify the parts of the input $x$ that contribute to the model prediction $y$. Because the attack is small and localized,  we can hope to recover the true class of input $x$ if we evaluate the model on a segmented input that contains no part of the attack. In the following, we look into the details of this phase. First, we present a segmentation-based approach to propose classes (see \Cref{fig:approach_overview}.a). Then, starting from proposed classes and the given input, we generate a mask for $x$ that may contain the malicious object (see \Cref{fig:approach_overview}.b).

\subsubsection{Class Proposal via Segmentation}
The detection of the attack begins with the identification of a set of possible classes that may be predicted by the model $f_m$. The first of such classes is the actual prediction, i.e., $y = f_m(x)$. The other classes are identified by segmenting the input $x$ and then evaluating the network on each segment. \Cref{alg:class_proposal} shows the algorithm to propose classes via input segmentation. 
Different approaches can be used to segment a given input $x$ including sliding windows and network-based region proposals~\cite{ren2015fasterrcnn}.
In our approach, we use the selective search image segmentation algorithm~\cite{uijlings2013selectivesearch}. Selective search generates an exhaustive list of region proposals based on the patterns and edges found in natural scenes~\cite{uijlings2013selectivesearch}. Then, we evaluate each proposed segment, i.e., $f(x_p)$, and return the $k$ most confident predictions, where $k$ is a configuration parameter of \tool{}.  We exclude the primary class $y=f_m(x)$ from our choice of $k$ classes.
A general guideline towards selecting $k$ is to set it to be slightly higher than the amount of classes that are present in the dataset per-image.  
In our case, Imagenet~\cite{deng09imagenet} has around 1.5 classes per image, and we set $k$ to 2.


\begin{algorithm}
    \footnotesize
    \Input{$f_m$ -- model\;
    ~~~~~~$x$ -- input of $f_m$\;
    ~~~~~~$y$ -- primary class, i.e., $y=f_m(x)$\;
    ~~~~~~$k$ -- propositions}
    \Output{$C$ - set of proposed classes and confidence ($|C|=k$)}

    $P$ = \SelectiveSearch{$x$}\;
    $C$ = $\{(y_p, conf_p): \forall x_{p} \in P, (y_p, conf_p) = f_m(x_{p})\wedge y_p \neq y\}$\;
    $C$ = \TopConf{$C$, $k$}\;
    \Return{$C$}

    \caption{ClassProposal}
    \label{alg:class_proposal}
\end{algorithm}

\subsubsection{Mask Generation}
Once the class proposal C is obtained, the second step of \tool{} consists of identifying the regions of $x$ that most highly influence the predictions C. To find these regions, we resort to techniques to explain and interpret model predictions. 



A particularly suitable approach for our goal is Grad-CAM~\cite{chattopadhyay2017gradcamplus}, a model-interpretation technique that identifies contiguous spatial regions of an input without requiring modifications to the original model. At a high level, Grad-CAM uses gradients computed in the final layers of a network to calculate the saliency of input regions. For class $c$, Grad-CAM calculates the gradients of the model's output $y^c$ (the model's \emph{logit} score for class $c$) with respect to each of the $k$ feature maps $A^k$ of the model's final pooling layer to obtain $\frac{\delta y^c}{\delta A^k}$. The mean gradient value of each filter map, or ``neuron importance weight'', is denoted $\alpha_c^{k} \coloneqq \frac{1}{Z} \Sigma_i \Sigma_j \frac{\delta y^c}{\delta A^k}$. Finally, the feature maps $A^k$ are weighted by their neuron importance and aggregated to obtain the final Grad-CAM output: $L^c_{Grad-CAM} \coloneqq ReLU(\Sigma_k \alpha_c^{k} A^k)$. Here, $ReLU(x)=\max(x, 0)$ is the ReLU activation function~\cite{glorot2011deeprelu} which retains only the positive gradient signals for class $c$. The output of Grad-CAM is a coarse heatmap of the positive importance of the image, usually at a lower resolution that the input image due to downsampling in the model's convolutional and pooling layers. Finally, masks are produced by binarizing the heatmap with a threshold of 15\% of max intensity.  We use this mask to segment salient regions for the next steps.

\begin{figure}[ht]
    \centering
	\begin{subfigure}[b]{1\columnwidth}
        \includegraphics[width=1\columnwidth]{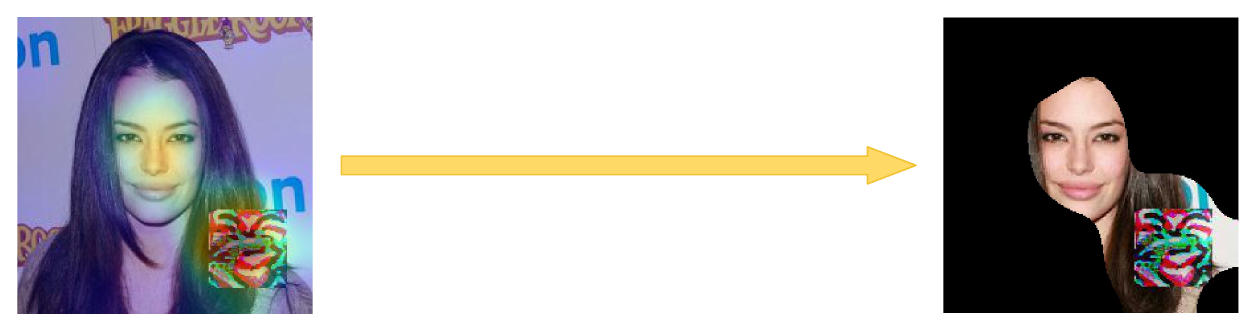}
        \caption{}
    \end{subfigure}
    
	\begin{subfigure}[b]{1\columnwidth}
        \includegraphics[width=1\columnwidth]{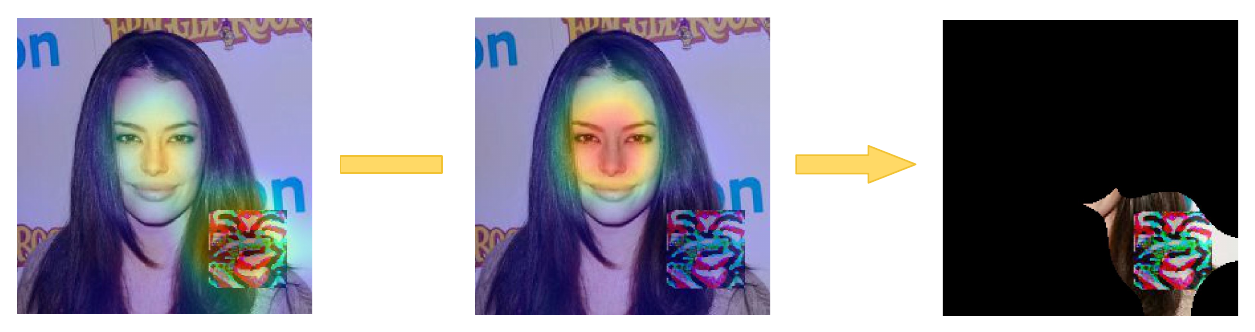}
        \caption{}
    \end{subfigure}
    \caption{Differential mask generation to remove ambiguous areas. \textbf{(a)} Example of Grad-CAM mask using prediction $y$ covering benign areas. \textbf{(b)} Example of precise mask by removing masks of class proposals $C$.}
    \label{fig:precise_mask}
\end{figure}




We generate masks with Grad-CAM as shown in \Cref{alg:mask_generation}. We start by extracting the mask using Grad-CAM for the input $x$ and prediction $y$. As Grad-CAM can identify salient areas for benign classes, the resulting mask may span over both benign and malicious regions. An illustrative example is in \Cref{fig:precise_mask}, where the Grad-CAM generated heatmap for a facial recognition network covers both a trojan trigger patch but also the original face. To improve the accuracy of the mask, we also extract a mask for each proposed class $y_{p}$. Then, we use these additional heatmaps to generate secondary masks to improve the original mask for the prediction $y$ by subtracting common regions. This results in a set of masks that highlight only the localized attack and not the other salient regions in the image.

\begin{algorithm}
    \footnotesize
    \Input{$f_m$ -- model\;
    ~~~~~~$x$ -- input for $f_m$\;
    ~~~~~~$y$, $conf$ -- model prediction on $x$\;
    ~~~~~~$C$ -- proposed classes}
    \Output{$M$ - masks for candidate regions with the malicious object}

    $mask_y$ = \MaskGradCAM{$f_m$, $x$, $y$}\;
    $M$ = $\{ (mask_y-$\MaskGradCAM{$f_m$, $x$, $y_{p}$}$, conf_{p})$ 
            $: (y_{p}, conf_{p}) \in C \}$\;
    \Return{$\{mask_y\} \cup M $}

    \caption{MaskGeneration}
    \label{alg:mask_generation}
\end{algorithm}

\subsection{Attack Detection}
\label{sec:fuzzing}

Once we identified regions $M$ of the inputs $x$ that may be containing malicious inputs, we test the model $f_m$ to determine whether any of the regions can hijack the expected predictions of a set of benign test inputs. 


\subsubsection{Test Generation} Once an input region is localized, \tool{} observes the effects the region has on the model to determine whether the region is adversarial or benign. To do so, \tool{} overlays the suspected region on a set of benign test images $X$, which are often shipped together with deployed models. Test images $X$ are fed back into the network, where the number of fooled examples are counted and used for adversarial images. Intuitively, the higher the number of mutated images can fool the model, the more likely the suspected region is an adversarial attack.


When the recovered region is small, this technique is effective at distinguishing adversarial and benign inputs, as small benign objects cannot typically overwhelm a network’s prediction. However, one problem of this approach is that a region (whether adversarial or benign) that covers a larger fraction of the input image will likely cause misclassifications when overlaid onto most other images because they occlude the original object.

To measure the extent to which an overlaid region causes misclassifications simply by occluding the original object, rather than by virtue of highly-salient features, we perform a second test where we replace the content of the overlaid region with an inert pattern of low saliency (e.g., Gaussian noise). We expect that adversarial regions will cause many misclassifications when overlaid on the test set, but have little effect on the network when replaced by an inert pattern before being overlaid. In contrast, we expect benign patterns to either cause few misclassifications, or to occlude objects and thus also disrupt the model when replaced by inert patterns. Algorithm~\ref{alg:testing} shows the algorithm we use to test extracted regions. For each region, we check how many overlaid test images fooled the network, and how confident the model is when classifying images overlaid with inert patterns.

\begin{algorithm}

    \Input{$f_m$ -- model\;
    ~~~~~~$x$ -- input for $f_m$\;
    ~~~~~~$y$ -- class of $x$\;
    ~~~~~~$M$ -- proposed masks\;
    ~~~~~~$X$ -- bening test images\;}
    \Output{$Fooled$, $AvgConf$}

    $R$ = $\{ x * mask : mask \in M\}$\;
    $IP$ = \InertPattern($M$)\;
   
    $X_{R}$ = \Overlay($X$, $R$)\;
    $X_{IP}$ = \Overlay($X$, $IP$)\;

    $fooled_{y_R}$ = $0$, $avg_{{conf}_{IP}}$ = $0$\;

    \For{$x_{R}$,$x_{IP} \in X_{R}, X_{IP}$}{
        $(y_R, conf_R), (y_{IP}, conf_{IP})$ = $f(x_{R}), f(x_{IP})$\;

        \If{$y_R == y$}{
            $fooled_{y_R}$ += 1
        }
        
        $avg_{{conf}_{IP}}$ += $conf_{IP}$
    }
    $avg_{{conf}_{IP}}$ = $\frac{avg_{{conf}_{IP}}}{|X|}$\;
    
    \Return $fooled_{y_R}$, $avg_{{conf}_{IP}}$\;
    \caption{Testing}
    \label{alg:testing}
\end{algorithm}

\subsubsection{Decision Boundary for Detection} 

With these two met- rics (number of images fooled and average inert pattern confidence values) we can determine whether an input x is adversarial. A naive approach is to use thresholding based rules, but it is hard to determine how to set the thresholds and which metric holds more importance. Instead, we use metrics collected on clean examples to train a simple two-feature one- class classifier and classify outliers as adversarial.

\begin{figure}[ht]
        \centering
        \includegraphics[width=1.0\columnwidth]{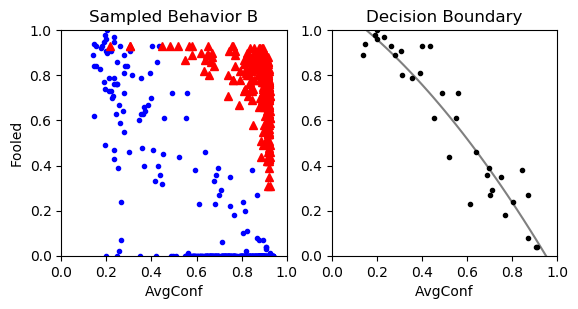}
        \caption{Example of boundary detection for the trojaned experiments using 400 data points and random noise as inert pattern.  On the left, the adversarial and benign metrics are plotted as red triangles and blue circles respectively; on the right, we plot the curve proposal from the sampled points, i.e., the benign test images $X$.}
    \label{fig:example_boundary detection}
\end{figure}

We examine this problem by plotting metrics from an example task in a 2D plot in Figure \ref{fig:example_boundary detection}, where the red triangular dots represent adversarial examples and the blue circular dots represent clean examples. We observe that the adversarial and clean points can be easily separated using a parabolic function, suggesting that a classifier approach such as linear regression or support vector machine could be used. However, this would operate under the false assumption that we have prior examples of adversarial inputs. Ideally, we want to create a technique that would allow us to identify an unseen adversarial input as an attack based on the attack-agnostic metrics.


We notice a general pattern where adversarial examples are usually clustered near the top-right of the plot.  Because adversarial examples are designed to cause misclassifications when applied to other images and not overly obstruct the other important regions of an image, it makes sense that both the number of images fooled and the average noisy confidence metrics of an adversarial example would be high. We can use our set of benign test images to generate statistics of the metrics of normal examples, which will then allow us to formulate when an input is abnormally adversarial.

We can use our collected metrics on clean examples to approximate a curve. By taking the points most likely to be the bounds of the statistics of our examples, we can approximate a curve where points lying outside our curve function can be classified as attacks. We collect our points by taking the points with the highest y-values for x-intervals, and then using a non-linear least squares function to approximate our curve. We then use our approximated curve to set classify attacks by calculating the distance between a curve and a point ---using the Constrained Optimization by Linear Approximation (COBYLA) method~\cite{conn1997cobyla}---and determining whether that distance is within a threshold estimated by the distances of clean examples lying outside the curve.

\begin{algorithm}[ht]
   
    \Input{$B$ -- sampled behavior of $f_m$}
    \Output{$f_{curve}$ -- approximated curve function\; 
    ~~~~~~~$d$ -- acceptable distance from $f_{curve}$}

    $f_{curve}$ = \ApproximateCurve{\OutPts{$B$}}\;
    
    $avg_d$ = $0$\;
    \For{$(x, y) \in B$}{
        \If{$f_{curve}(x) > y$}{ 
            $avg_d$ += \COBYLA{$(y, x), f_{curve}$}\;
        }
    } 
    $d$ = $\frac{avg_d}{|B|}$\;
    \Return{$f_{curve}$, $d$}\;

    \caption{DecisionBoundary}
    \label{alg:decision_boundary}
\end{algorithm}

\section{Evaluation}\label{sec:evaluation}


In this section, we evaluate \tool{} on a variety of previously proposed attacks, to highlight its versatility. In Section~\ref{sec:adaptive_attacks}, we then perform an extensive evaluation against \emph{adaptive} adversaries with full knowledge of \tool{}'s inner workings---thereby demonstrating \tool{}'s robustness.

After presenting the experiment setup in \Cref{sec:parameters}, in \Cref{sec:known_attacks}, we evaluate the effectiveness of \tool{} in protecting networks from three known attacks, i.e., adversarial patches, trojan triggers, and backdoors, with an average true positive rate of 96.22\% and an average true negative rate of 95.36\%. We further evaluate \tool{} in a realistic physical setting with a printed adversarial patch (see Section~\ref{sec:physical}).

Finally, in \Cref{sec:performance}, we look into the latency of \tool{} in terms of runtime analysis.



\subsection{General Experiment Settings}
\label{sec:parameters}

We evaluated \tool{} when protecting three publicly available networks shared from prior works. Two of the selected networks are compromised and one is uncompromised. The compromised networks are a backdoored Faster-RCNN network for reading signs detection by Gu et al.~\cite{gu2017badnets} and a VGG-16 trojaned network for facial recognition by Liu et al.~\cite{Liu2018TrojaningAO}. The uncompromised network is a VGG-16 network trained on the Imagenet dataset by Simonyan et al.~\cite{simonyan2018vgg16}. Additionally, to operate, \tool{} requires a benign test image set $X$ and an inert pattern $s$ to generate the decision boundary as shown in \Cref{sec:fuzzing}. We present the generation of each test set $X$ for the selected networks in \Cref{sec:known_attacks}. Unless specified, we use random noise for our inert pattern $s$.

We measure the effectiveness and robustness of \tool{} in terms of accuracy and performance, collecting the TP/TN and FP/FN rates of \tool{} when processing benign and adversarial inputs. We present the creation of benign and adversarial images datasets of our experiments in \Cref{sec:known_attacks}. We measure efficiency by evaluating the execution of each step of \tool{}.

Finally, 
\tool{} uses Tensorflow 1.5 to generate the adversarial patches for the uncompromised network, BLVC-Caffe for the trojaned network, and Faster-RCNN Caffe~\cite{girshick2015fastrcnn} for the poisoned network. To parallelize the class proposal, \tool{} relies on the ROI pooling layer as implemented by the Fast-RCNN Caffe version. Finally, \tool{} uses the off-the-shelf implementations of Grad-CAM~\cite{selvaraju2016gradcam} and selective search~\cite{uijlings2013selectivesearch}.

\subsection{Effectiveness}
\label{sec:known_attacks}

The first part of our evaluation assesses the effectiveness of \tool{} in protecting our selected networks against three attacks, i.e., trojan triggers~\cite{Liu2018TrojaningAO} against trojaned networks, backdoors~\cite{gu2017badnets} against poisoned networks, and adversarial patches~\cite{brown2017adversarialpatch} against uncompromised models. For each attack, we measured the effectiveness of \tool{} in terms of the number of attacks detected. For the adversarial patches attack, we considered an additional variant where the attacker uses multiple patches at the same time. The summary of this evaluation is in \Cref{tab:known_attacks_results}.

\begin{table*}
	\centering
	\caption{Effectiveness of \tool{} protecting compromised and uncompromised networks.}
	\begin{tabular}{l l l c r r r r}
		\toprule
		\textbf{Network and Task}                                 & \textbf{Attack}            & \textbf{Inert Noise} & \multicolumn{1}{c}{\textbf{TP}} & \multicolumn{1}{c}{\textbf{TN}} & \multicolumn{1}{c}{\textbf{FP}} & \multicolumn{1}{c}{\textbf{FN}} \\

		\midrule
		VGG-16~\cite{Liu2018TrojaningAO} for face recognition     & Trojan trigger             & Random               & 99.2\%                          & 99.8\%                          & 0.3\%                           & 0.8\%                           \\
		VGG-16~\cite{Liu2018TrojaningAO} for face recognition     & Trojan trigger             & Checker              & 99.2\%                          & 99.5\%                          & 0.5\%                           & 0.8\%                           \\
		Faster-RCNN~\cite{gu2017badnets} for road signs detection & Backdoor                   & Random               & 85.0\%                          & 86.9\%                          & 13.1\%                          & 14.8\%                          \\
		VGG-16~\cite{simonyan2018vgg16} for ImageNet              & Adversarial patch          & Random               & 98.5\%                          & 95.3\%                          & 4.8\%                           & 1.5\%                           \\
		VGG-16~\cite{simonyan2018vgg16} for ImageNet              & 2$\times$Adversarial patch & Random               & 99.2\%                          & 95.3\%                          & 4.8\%                           & 0.8\%                           \\
		VGG-16~\cite{simonyan2018vgg16} for ImageNet              & Physical adversarial patch & Random               & 93.7\%               & 90.2\%                & 9.8\%                & 6.3\%                \\
		\bottomrule
	\end{tabular}
	\label{tab:known_attacks_results}
\end{table*}


\subsubsection{Trojaned Networks}

A trojaned network is a model whose weights have been modified by an adversary to implant a response when processing inputs containing a so-called \emph{trigger trojan}~\cite{Liu2018TrojaningAO}. The trigger trojan is created from a set of specified neurons, and then fine-tunes the model's weights on the trigger to hijack the prediction~\cite{Liu2018TrojaningAO}. For our evaluation, we selected the facial recognition model shared by Liu et al.~\cite{Liu2018TrojaningAO} with a VGG-16 architecture~\cite{simonyan2018vgg16} trained on the VGG-Face dataset~\cite{parkhi2015vggface}.

\begin{figure}[t]
	\centering
	\includegraphics[width=0.48\columnwidth]{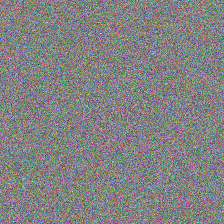}~
	\includegraphics[width=0.48\columnwidth]{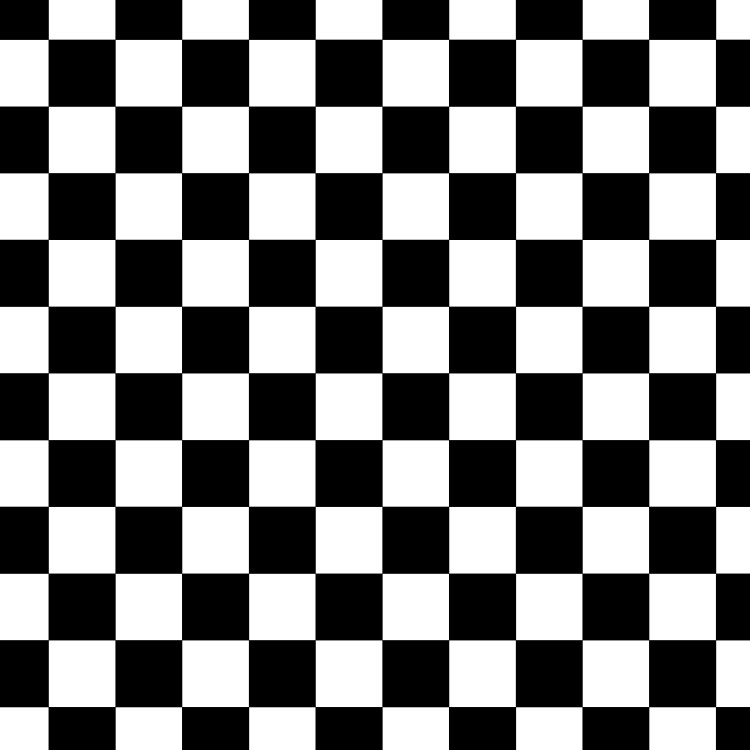} \\
	\caption{Examples of inert patterns. The default pattern we use for $s$ is random noise shown on the left.  Another pattern we can potentially use is the checkered pattern on the right which the VGG-Face network also responds weakly to.}
	\label{fig:secret_patterns}
\end{figure}

\begin{figure*}
	\centering
	\begin{subfigure}[b]{.234\linewidth}
		\centering
		\includegraphics[width=0.48\columnwidth]{figures/trojan_cropped.png}
		\includegraphics[width=0.48\columnwidth]{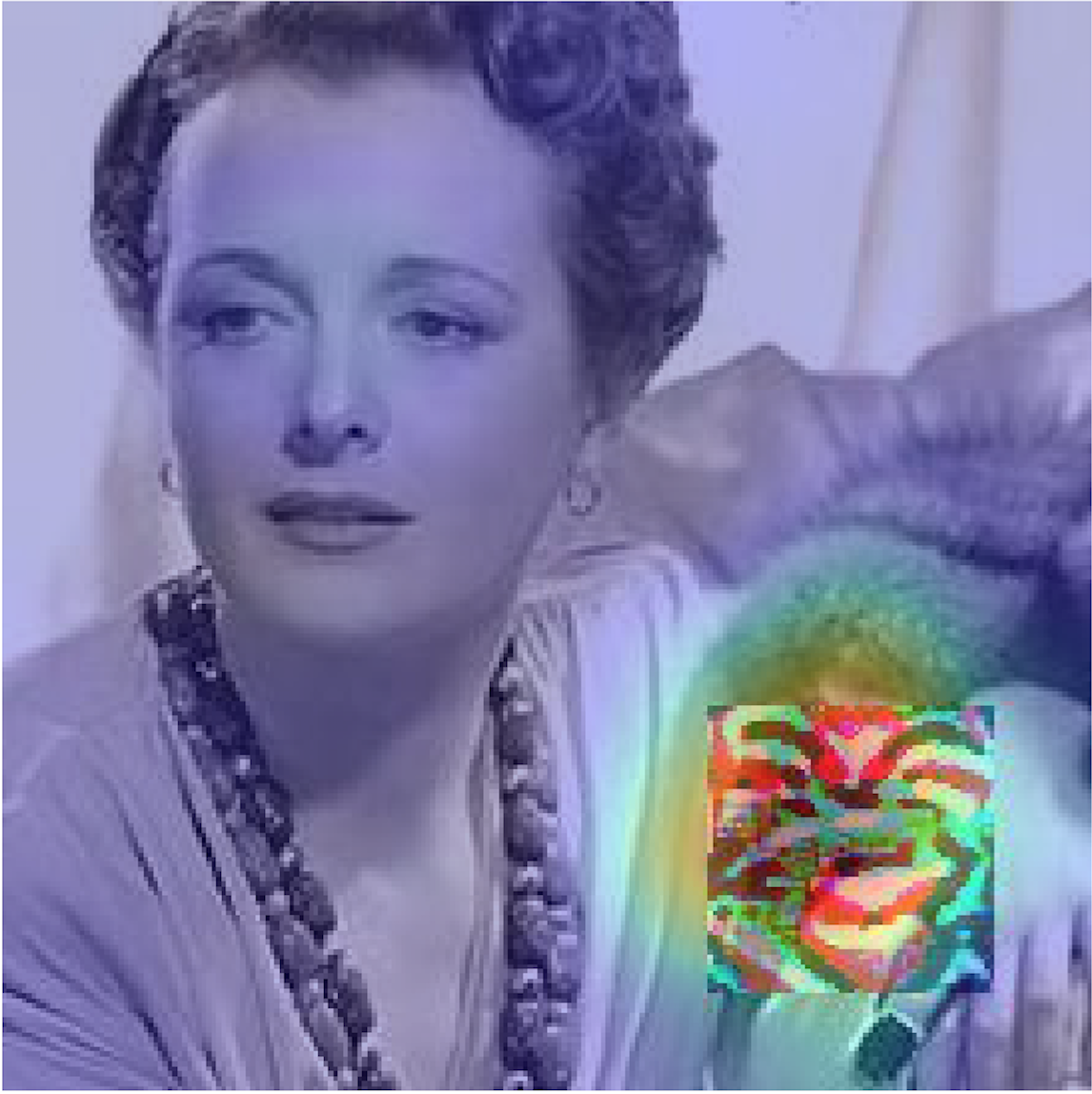}\\
		\vspace{0.45cm}
		\includegraphics[width=1\columnwidth]{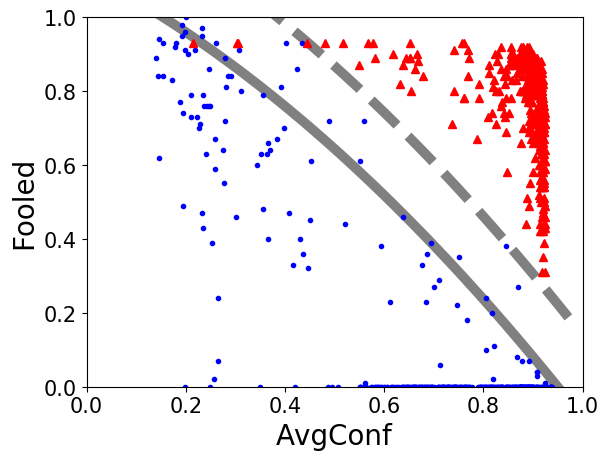}
		\caption{Trojan trigger.}
		\label{fig:trojan_qualitative}
	\end{subfigure}\hfill
	\begin{subfigure}[b]{.235\linewidth}
		\centering
		\includegraphics[width=0.48\columnwidth]{figures/flower_nobg.png}
		\includegraphics[width=0.43\columnwidth]{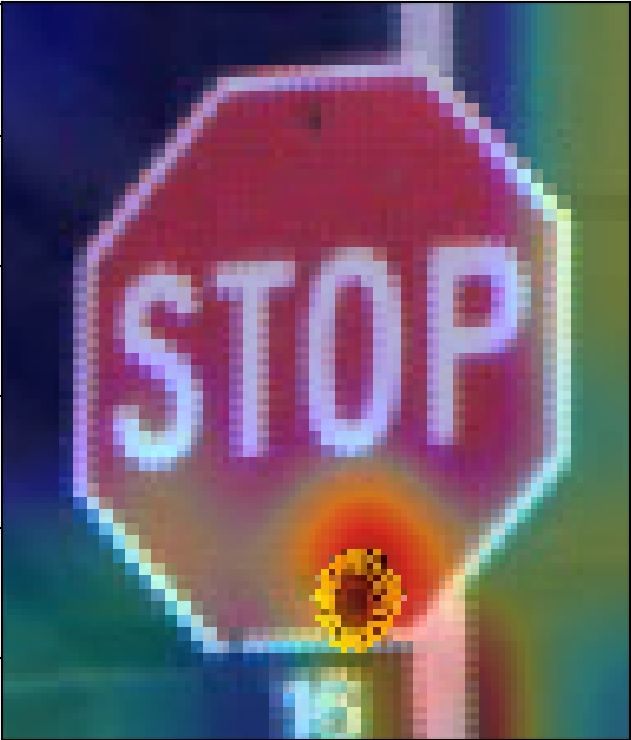}\\
		\vspace{0.45cm}
		\includegraphics[width=1\columnwidth]{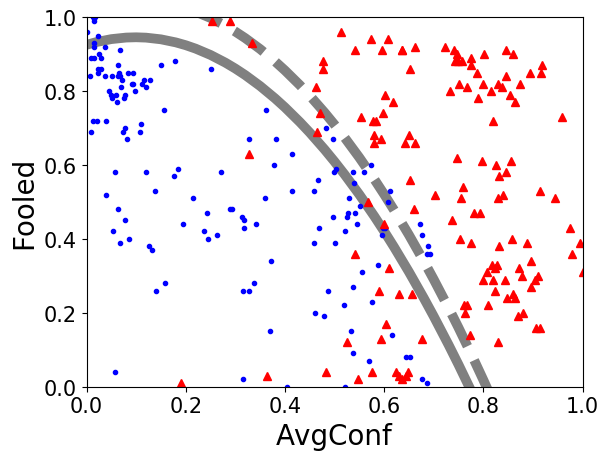}
		\caption{Backdoor.}
		\label{fig:badnets_qualitative}
	\end{subfigure}\hfill
	\begin{subfigure}[b]{.234\linewidth}
		\centering
		\includegraphics[width=0.48\columnwidth]{figures/adversarial_patch.png}
		\includegraphics[width=0.48\columnwidth]{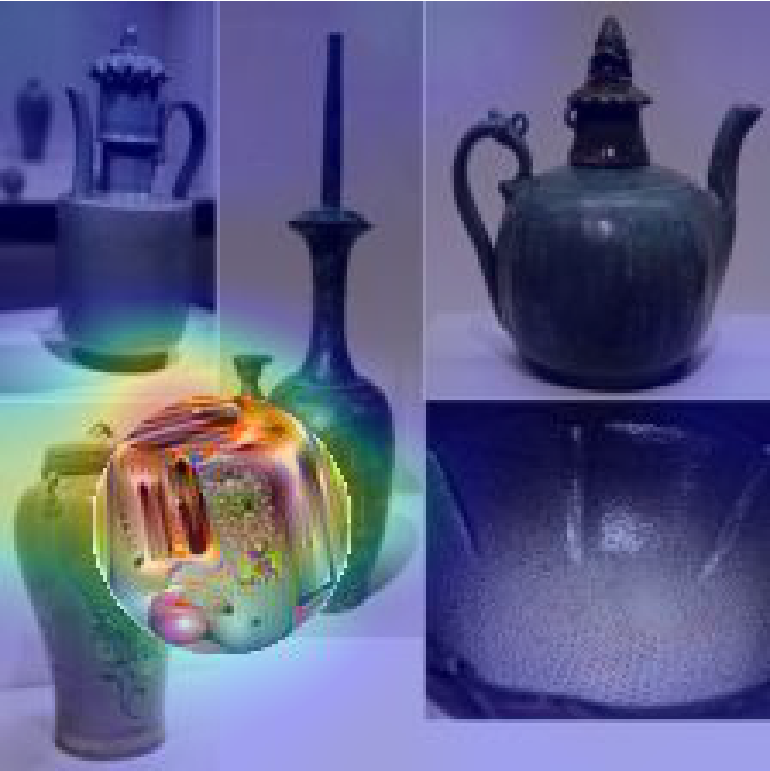}\\
		\vspace{0.45cm}
		\includegraphics[width=1\columnwidth]{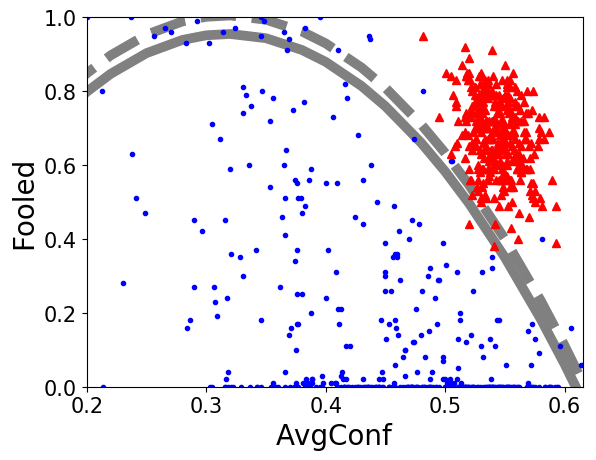}
		\caption{Adversarial patch.}
		\label{fig:adversarialpatch_qualitative}
	\end{subfigure}\hfill
	\begin{subfigure}[b]{.234\linewidth}
		\centering
		\includegraphics[width=0.48\columnwidth]{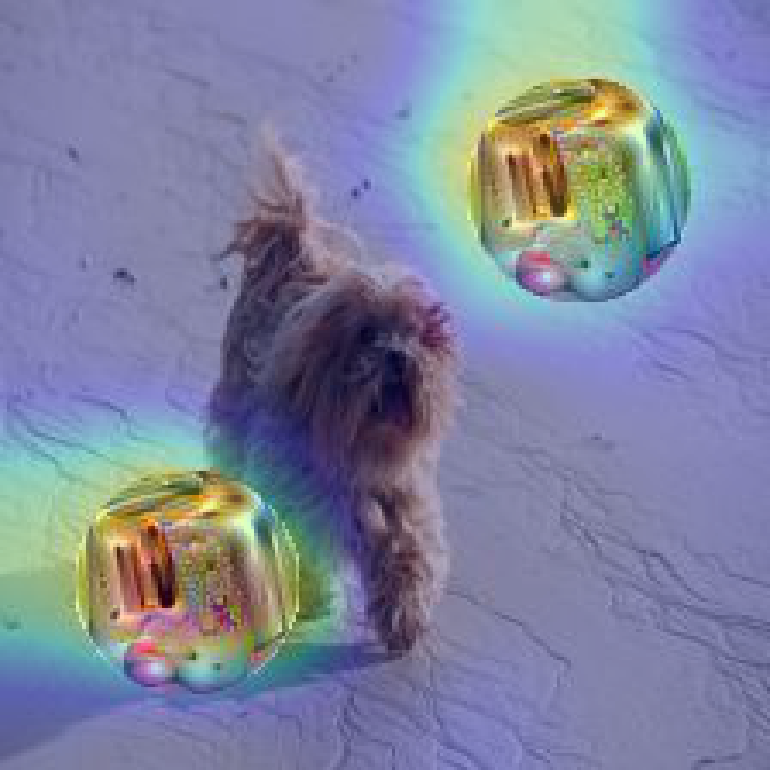}
		\includegraphics[width=0.48\columnwidth]{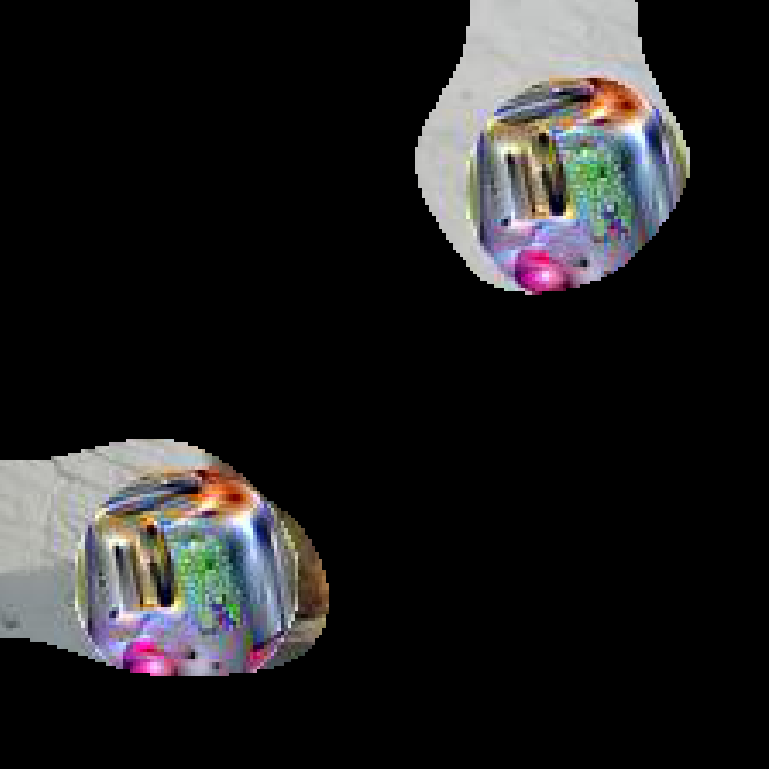}\\
		\vspace{0.45cm}
		\includegraphics[width=1\columnwidth]{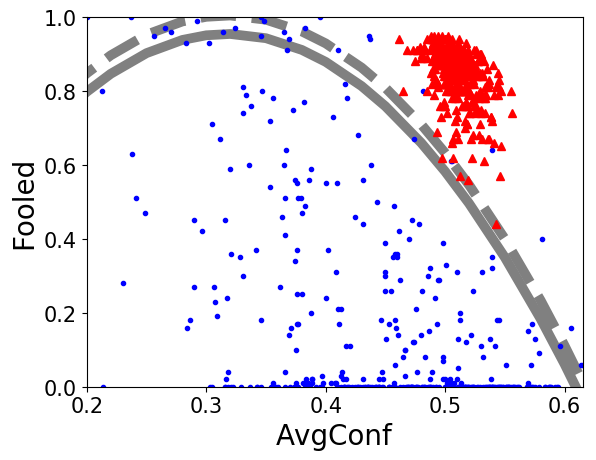}
		\caption{2$\times$Adversarial patch.}
		\label{fig:double_qualitative}
	\end{subfigure}
	\caption{Evaluation inputs and results. The first row shows examples of adversarial objects used by the attacker and the Grad-CAM output. The bottom row shows scattered plots of benign inputs (in blue) and adversarial inputs (in red) with respect to the decision boundaries (solid lines) and thresholds (dashed lines) used by \tool{} for the attack detection.}
	\label{fig:evaluation_results}
\end{figure*}

\paragraph{Datasets} For the evaluation, we created three datasets with benign, adversarial, and test images as follows. The benign images dataset contains 400 images randomly selected from the LFW dataset~\cite{huang2007lfw}. From the benign dataset, we derived the adversarial inputs dataset by placing the square trigger trojan shared by Liu et al.~\cite{Liu2018TrojaningAO} to hijack the prediction of the VGG-16 network to the class ``A.J. Buckley''. As the trigger trojan is created from a fixed region of the input, we placed them in the same position, i.e., the bottom-right corner of each image.  An example of such a trigger trojan and adversarial input is shown in \Cref{fig:trojan_qualitative}. Then, the test set $X$ contains 100 randomly selected images from the VGG-16 dataset~\cite{simonyan2018vgg16}. Finally, in this evaluation, we considered two types of inert noise, i.e., random noise and checker pattern (Examples of the two types of inert noise are in \Cref{fig:secret_patterns}).

\paragraph{Results} Of the 400 adversarial images, 368 of them hijack the prediction of the compromised model, i.e., the model incorrectly predicted ``A.J. Buckley''. Of these 368 attacks, \tool{} can successfully detect almost all of them. i.e., 365 malicious inputs resulting in TP rates above 99\%, regardless of the used inert noise. Then, the number of TN is also considerably high, with rates above 99\%. In this case, the inert noise has a very marginal role in the detection. Of the 400 benign images, we have one FN when using the random inert noise and two FNs when using the checker inert noise.

To better understand the behavior of \tool{}, we have a closer look at the distribution of the benign and adversarial images with respect to the decision boundary used by \tool{}. The plot is shown in \Cref{fig:trojan_qualitative}. As expected, most of the benign samples (in blue) fail to fool the model except for few outliers which produce low $avgConf$ scores. On the contrary, the adversarial inputs (in red) fool the model a high number of times with a very high $avgConf$ confidence scores. The decision boundary is calculated using the test set $X$ as explained in \Cref{sec:fuzzing}, and it is well separating the adversarial from benign regions.

\subsubsection{Poisoned Networks}

A poisoned network is a model created by an adversary via training, with the goal to implant a specific model response when processing inputs containing a \emph{backdoor}. As opposed to trojaned networks, poisoned networks are compromised \emph{during} the training phase instead of after. For our evaluation, we selected the poisoned network shared by Gu et al.~\cite{gu2017badnets}, a Faster-RCNN model~\cite{ren2015fasterrcnn} to detect stop signs. The authors poisoned the training set so that the network will incorrectly classify stop signs with a yellow flower---the backdoor---as a warning sign. The backdoor and an example of adversarial input is shown in \Cref{fig:badnets_qualitative}.

\paragraph{Datasets} For this evaluation, the benign images dataset contains 145 images of stop signs from the LISA dataset~\cite{mogelmose2014traffic} by cropping out the images with their labeled bounding boxes. Then, we derived the adversarial images by placing the yellow flower backdoor shared by Gu et al.~\cite{gu2017badnets} at approximately below the ``stop'' text of each sign to create an ``attack'' dataset. The test set $X$ contains 100 images selected from each class with equal probability.


\paragraph{Results} Out of the 145 adversarial inputs, 134 images successfully fool the Faster-RCNN model. \tool{} detected 114 images, with a TP rate of 85\%. When using benign images, \tool{} incorrectly detected 19 images of 145 benign inputs, resulting in a 14.8\% FNs. While \tool{} can achieve much higher accuracy and precision, we note that \tool{}'s accuracy in this experiment is comparable to the accuracy of proposed adversarial detection techniques (see, e.g.,~\cite{carlini2018eval}).

Higher FP and FN rates of \tool{} can be better explained by having a look at the distribution of benign and adversarial images in \Cref{fig:badnets_qualitative}. As opposed to trojaned networks, benign and adversarial images are not densely grouped. Instead, they are rather scattered over large areas, crossing the decision boundary in two points, i.e., benign images above the boundary decision in the center of the plot and adversarial inputs below the boundary decision. Scattered points as shown in \Cref{fig:badnets_qualitative} are likely due to coarser heatmaps generated by Grad-CAM when applied to the Faster-RCNN network. The Faster-RCNN architecture uses a 12 $\times$ 12 ROI pooling layer, whose layer is used by Grad-CAM to create the heatmap. The ROI pooling layer of the Faster-RCNN is smaller than the other networks used in our evaluation. For example, the VGG-16 trojaned network has a 16$\times$16 ROI pooling layer. A smaller ROI pooling layer results in coarser grained heatmaps, whose masks are likely to include regions unrelated to the targeted class. As a result, test images with overlaid adversarial regions will include ambiguous features affecting both the probability of fooling the network and the confidence of the classification.

\subsubsection{Uncompromised Networks}
\label{sec:uncompr_networks_eval}

In this section, we evaluate \tool{} protecting an uncompromised network when processing adversarial patches. In our case, we use a VGG-16~\cite{simonyan2018vgg16} Imagenet-pretrained~\cite{deng09imagenet} network as our target network $f_m$.

\paragraph{Datasets} We randomly sampled 400 images from the Imagenet test set~\cite{deng09imagenet} to implement our attack. Then, we placed the patch at random positions on the 400 images to generate our dataset of adversarial images, occupying around 25\% of the input images. We used the adversarial patch created by Brown et al.~\cite{brown2017adversarialpatch} which can fool the Imagenet-trained VGG-16 model to classify inputs as a ``toaster''. The patch and an example of adversarial input is shown in \Cref{fig:adversarialpatch_qualitative}. Also, in this evaluation, we considered a variant of the attack with images containing multiple adversarial patches. Finally, our test set $X$ consists of 100 randomly selected images from the Imagenet training set.

\paragraph{Results} Our evaluation showed that 338 images of the 400 adversarial images with a single patch fool the Imagenet-trained VGG-16. \tool{} detected almost all of them, i.e., 333 out of 338 with a TP rate above 98\%. When using two patches, the adversary has a slightly higher advantage and can hijack the prediction 369 times over 400 attempts. Of these 369 attacks, \tool{} has a stronger response than single-patch attacks, detecting 368 attacks, with almost 100\% TP rate.

In \Cref{fig:adversarialpatch_qualitative,fig:double_qualitative}, we observe that the adversarial images (in red) points are skewed towards producing both higher $numFooled$ and $avgConf$ numbers. The classification rule generated by the decision boundary produces TN rates of ${\sim}$95\% as seen in \Cref{tab:known_attacks_results}. Overall, both TP and FP compare favorably to other literature focused on adversarial attack detection.

Finally, when using two patches simultaneously, Grad-CAM correctly identifies the two disjoint regions that contribute to the prediction. \Cref{fig:double_qualitative} shows both the heatmap and the mask generated by \tool{}. Additionally, we measured how often \tool{} can detect both patches by checking whether the generated masks are disjoint and cover substantial portions of both patches. We find that \tool{} discovers both patches 97.4\% of the time, which we consider to be sufficiently accurate relative to the TP and TN values.

\subsection{Physical Attacks}
\label{sec:physical}

Localized adversarial attacks have the distinct advantage of being deployable in real-world settings as a physical attack via stickers or patches, as compared to adversarial noise.  Several prior works~\cite{brown2017adversarialpatch, sharif2016glasses, evtimov2017robust, eykholt2018physical, athalye2017synthesizing} have demonstrated how adversarial noise can be effectively synthesized in the physical world to fool visual classifiers fed with video streams.
In this section, we experimentally validate that \tool{}'s defensive capabilities perform well in a realistic attack scenario where a physically printed patch is displayed next to a common object, with the goal of robustly and reliably subverting a classifier's predictions.

\begin{figure*}
	\centering
	\begin{subfigure}[b]{.19\linewidth}
		\centering
		\includegraphics[width=1\columnwidth]{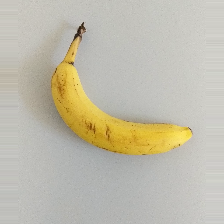}\\
		\vspace{0.12cm}
		\includegraphics[width=1\columnwidth]{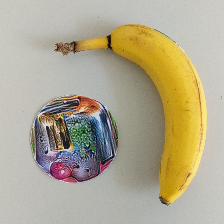}
	\end{subfigure}
	\begin{subfigure}[b]{.19\linewidth}
		\centering
		\includegraphics[width=1\columnwidth]{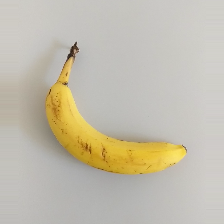}\\
		\vspace{0.12cm}
		\includegraphics[width=1\columnwidth]{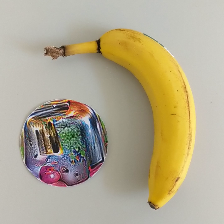}
	\end{subfigure}
	\begin{subfigure}[b]{.19\linewidth}
		\centering
		\includegraphics[width=1\columnwidth]{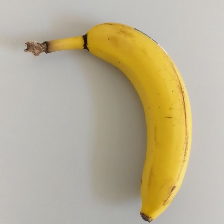}\\
		\vspace{0.12cm}
		\includegraphics[width=1\columnwidth]{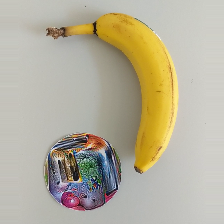}
	\end{subfigure}
	\begin{subfigure}[b]{.19\linewidth}
		\centering
		\includegraphics[width=1\columnwidth]{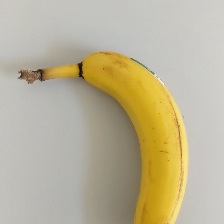}\\
		\vspace{0.12cm}
		\includegraphics[width=1\columnwidth]{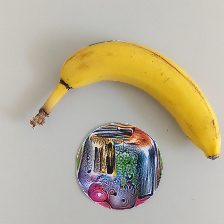}
	\end{subfigure}
	\begin{subfigure}[b]{.19\linewidth}
		\centering
		\includegraphics[width=1\columnwidth]{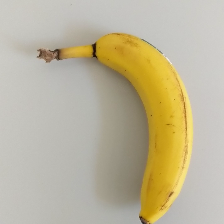}\\
		\vspace{0.12cm}
		\includegraphics[width=1\columnwidth]{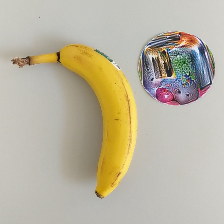}
	\end{subfigure}
	\caption{224x224 video frames without (top) and with (bottom) a physical adversarial patch used for our evaluation.}
	\label{fig:physical_atta}
\end{figure*}

We captured a video of a banana displayed on its own or next to an adversarial patch~\cite{brown2017adversarialpatch} with a variety of positions and orientations (see Figure~\ref{fig:physical_atta}). The undefended VGG16 network classifies all $610$ benign video frames correctly, and is always fooled by the patch when present in all other $541$ frames. 
We evaluate \tool{} on these video frames using the same parameters as in our previous (digital) adversarial patch experiments---including the use of the randomly selected Imagenet test set and the checkerboard pattern for fuzzing. 
With \tool{}, the patch is correctly identified in $93.7\%$ of the video frames, and the banana is correctly identified in $90.2\%$ of benign frames.

Albeit not perfect, these results are promising: with \tool{}, the adversarial patch fails its goal of reliably subverting the classifiers on the vast majority of frames. In a deployment scenario where a classifier averages predictions over multiple frames (e.g., by yielding an object prediction every $10$ frames, via a majority vote over the preceding frames), \tool{} successfully yields correct classifications for the full benign video, while detecting and rejecting the adversarially perturbed video.

\subsection{Runtime Analysis}
\label{sec:performance}

Finally, we measured the latency of \tool{} when protecting a VGG-16~\cite{simonyan2018vgg16} architecture as a case study. All components---except for the selective search algorithm---are executed on a Tesla K-80 GPU. A single execution of \tool{} consists of a run of the selective search algorithm, the Grad-CAM, and a forward pass with a batch of test images. Using the off-the-shelf implementations, the runtime of selective search and Grad-CAM is $2$ and $0.4$ seconds respectively. The forward pass through the network takes about $2.5$ seconds. However, we note that the forward pass runtime is greatly penalized by the overhead introduced by our PyCaffe implementation---a barebone forward pass of the VGG-16 is expected to take from $0.04$ to $0.1$ seconds~\cite{cnn_benchmarks}. With an expected worst-case execution runtime of $0.1$ seconds, and that the forward pass runtime is independent of the batch size, the total latency introduced by \tool{} amounts to $2.5$ seconds.


\section{Adaptive Attacks}
\label{sec:adaptive_attacks}

Our previous analysis demonstrates \tool{}'s versatility in protecting neural networks from a variety of attacks. In this section, we evaluate the more challenging and realistic threat model of a fully adaptive white-box adversary~\cite{carlini2019evaluating} that is aware of the presence of \tool{}, of its architecture and its inner workings. Such an adversary can attempt to create adversarial objects that simultaneously fool the target network while also compromising the core components of \tool{}, i.e., the mask generation, the class proposal, and the attack detection. In \Cref{sec:attacking_region_proposal}, we present attacks aimed at fooling the mask generation phase. Then, in \Cref{sec:attacking_class_proposal}, we evaluate attacks against the class proposal. Finally, in \Cref{sec:attacking_classification}, we present attacks against the attack detection phase.

\subsection{Attacking the Region Proposals}
\label{sec:attacking_region_proposal}

\begin{figure*}[ht]
	\centering
	\begin{subfigure}[b]{.24\linewidth}
		\centering
		\includegraphics[width=0.48\linewidth]{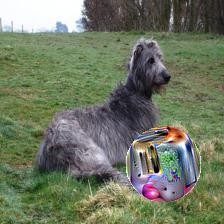}
		\includegraphics[width=0.48\linewidth]{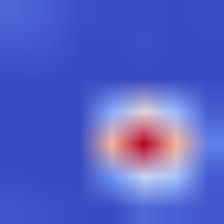}
		\caption{Heatmap of the adv. patch}
		\label{fig:dog_patch_1}
	\end{subfigure}~~
	\begin{subfigure}[b]{.24\linewidth}
		\centering
		\includegraphics[width=0.48\linewidth]{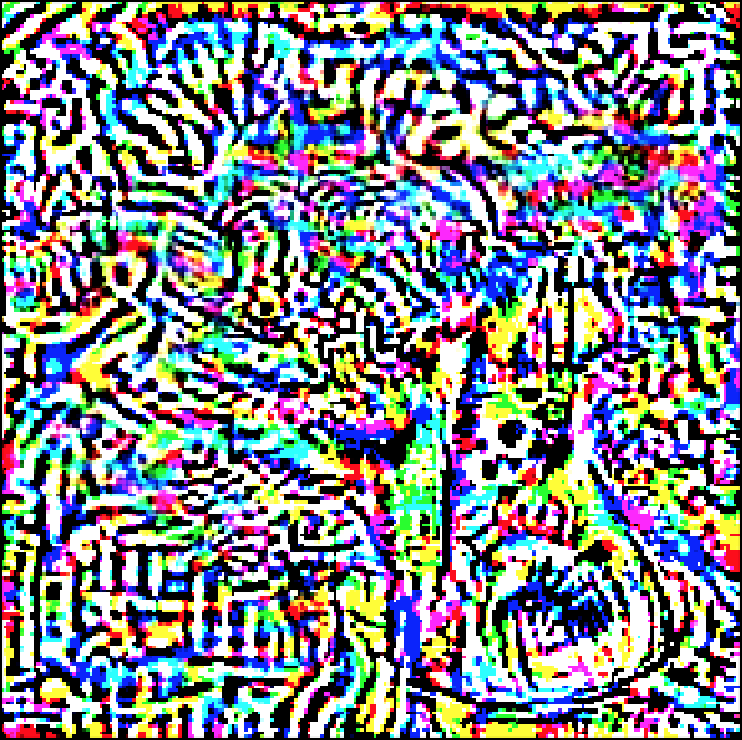}
		\includegraphics[width=0.48\linewidth]{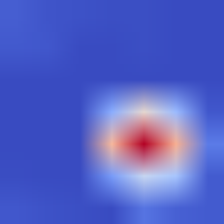}
		\caption{Full image perturbation}
		\label{fig:dog_patch_2}
	\end{subfigure}~~
	\begin{subfigure}[b]{.24\linewidth}
		\centering
		\includegraphics[width=0.48\linewidth]{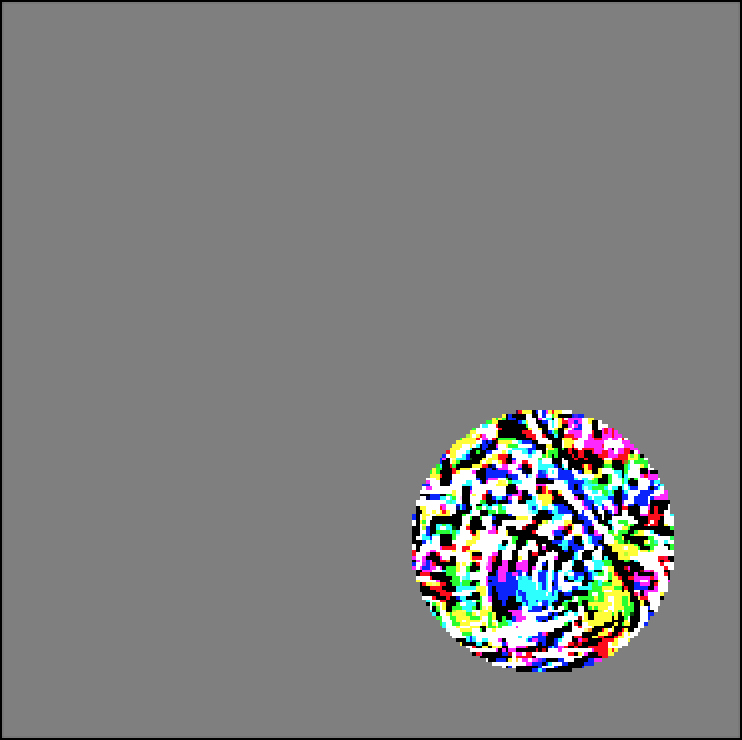}
		\includegraphics[width=0.48\linewidth]{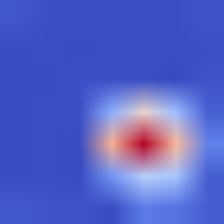}
		\caption{Localized perturbation}
		\label{fig:dog_patch_3}
	\end{subfigure}~~
	\begin{subfigure}[b]{.24\linewidth}
		\centering
		\includegraphics[width=0.48\linewidth]{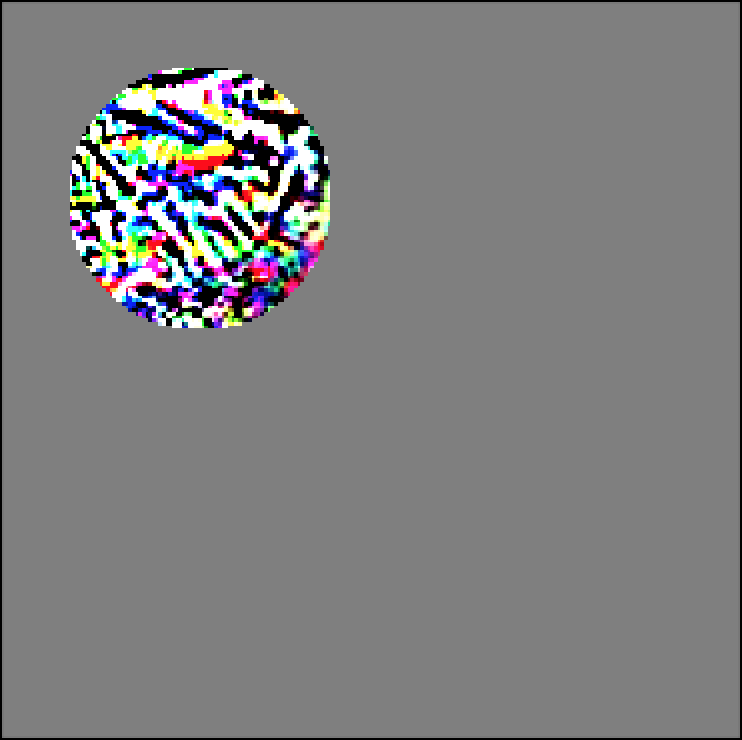}
		\includegraphics[width=0.48\linewidth]{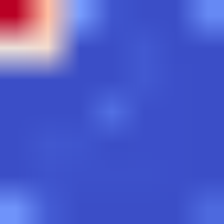}
		\caption{Misdirection}
		\label{fig:dog_patch_4}
	\end{subfigure}
	\caption{Attacking region proposal. \textbf{(a)} The heatmap of the adversarial patch overlaid on an image of a dog. \textbf{(b)} Full-image perturbations can generate a visually identical heatmap. \textbf{(c)} Localized perturbation can generate the same heatmap of (a). \textbf{(d)} Localized perturbation fails to generate a similar  heatmap.}
	\label{fig:dog_patch}
\end{figure*}

Our defense is reliant on successfully localizing the adversarial region in an image. In our current framework, this is done using the Grad-CAM algorithm, which generates a heatmap of the salient regions leading to a classification. If an attacker can disrupt the Grad-CAM mechanism and avoid successful detection and localization, the subsequent components of the pipeline will fail. We observe that the Grad-CAM mechanism uses network back-propagation to measure region importance. As this mechanism is differentiable, an adversary could generate targeted gradient perturbations to control the position and size of the heatmap. In this section, we consider three of these attacks: Grad-CAM perturbation, misdirection of the mask creation, and mask minimization. However, as the analysis of this section shows, while manipulating the heatmap is feasible in theory, in our defense context heatmap manipulation is ineffective in practice.


\subsubsection{Perturbing Grad-CAM}

Consider the adversarial image and its Grad-CAM heatmap in \Cref{fig:dog_patch_1} for the target class ``toaster''. The heatmap in \Cref{fig:dog_patch_1} correctly identifies the area containing the patch. An adversary could try to generate a perturbation, i.e., an adversarial image, resulting in a visually identical heatmap. Such a perturbation can be created in practice as the Grad-CAM function $L^c_{GradCAM} = ReLU(\Sigma_k \alpha_c^{k} A^k)$ is differentiable, and the adversary can optimize an input on this function given a target class. For example, the adversary can use a standard Stochastic Gradient Descent Optimizer (SGD) on a VGG-16 network~\cite{simonyan2018vgg16} to minimize a loss function calculated as the total difference between the current Grad-CAM output and the target Grad-CAM output, and iteratively add noise until the loss converges. An example of such a perturbation is shown in \Cref{fig:dog_patch_2}.

This attack shows that Grad-CAM outputs can be precisely manipulated through gradient optimization. However, such an attack requires to feed the network with a perturbation of the size of the input of the network being protected, which is not part of our threat model (\Cref{sec:threat_model}).




\subsubsection{Heatmap Misdirection}

In this attack, we consider an adversary who intends to create an input image that tricks \tool{} to detect the adversarial object in a different region where the adversarial object is located. Let us assume the adversary intends to convince \tool{} that the adversarial object is in the position as indicated by the heatmap of \Cref{fig:dog_patch_1}.
As we demonstrated earlier, an adversary allowed to add perturbational noise to an entire image can generate a heatmap at the desired position. As we discussed before, the adversary does not have full control of the whole image, rendering the attack infeasible. An attacker may try to generate a heatmap visually identical to the adversarial patch of \Cref{fig:dog_patch_1} by using a perturbation constrained to the same region. \Cref{fig:dog_patch_3} shows an example of such a perturbation obtained by modifying the loss function of the SGD to constrain the noise on the same region of interest. At this point, the adversary can try to generate a perturbation at a different location leaving the goal of the attack unchanged. An example of such an attempt is shown in \Cref{fig:dog_patch_4}. The perturbation is now on the top-left corner of the input image. However, the resulting heatmap is not shown in the desired position, i.e., bottom-right. Instead, the heatmap is positioned nearby the location of the perturbation, indicating that localized noise can only affect the corresponding Grad-CAM region.

\subsubsection{Heatmap Minimization}

\begin{figure}[htb]
	\centering
	\begin{subfigure}[b]{.49\linewidth}
		\centering
		\includegraphics[width=0.44\columnwidth]{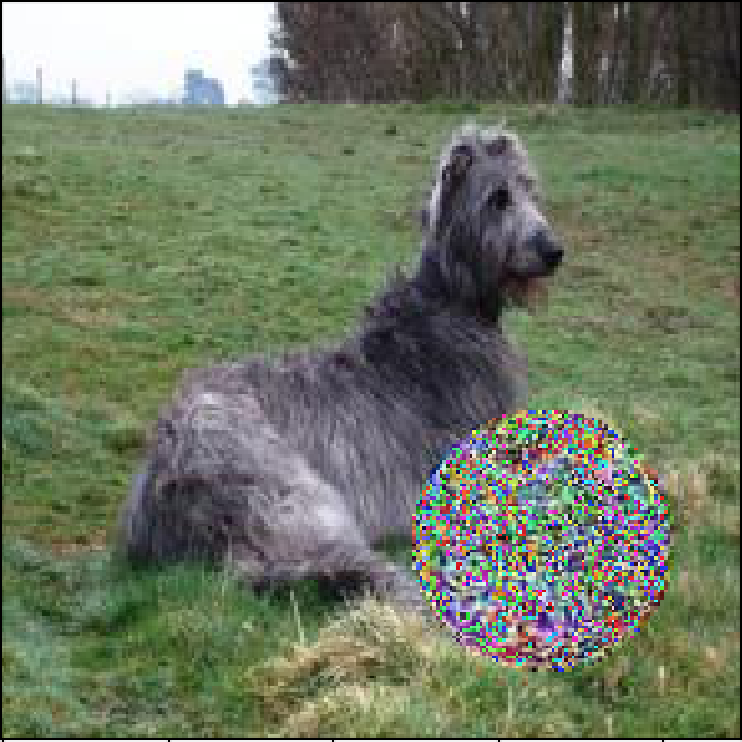}
		\includegraphics[width=0.44\columnwidth]{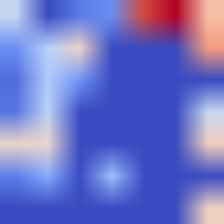}\\
		\includegraphics[width=1\columnwidth]{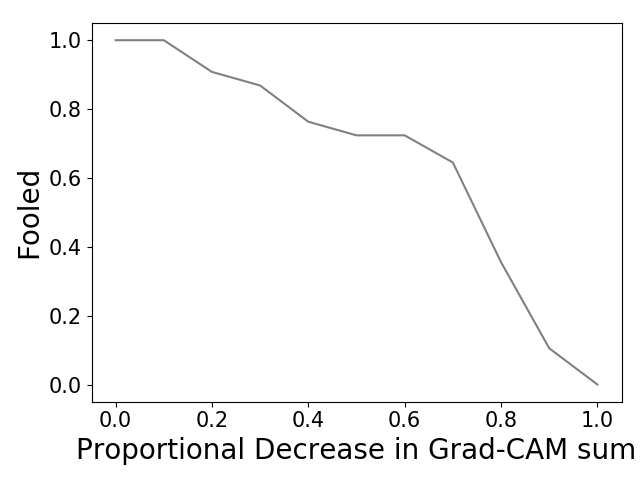}
		\caption{Heatmap minimization}
		\label{fig:gradcam_minimization}
	\end{subfigure}
	\begin{subfigure}[b]{.49\linewidth}
		\centering
		\includegraphics[width=0.44\columnwidth]{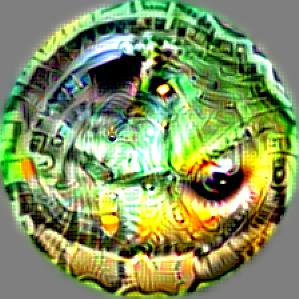}
		\includegraphics[width=0.44\columnwidth]{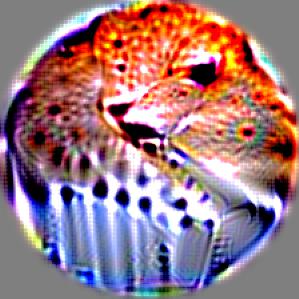} \\
		\includegraphics[width=1\columnwidth]{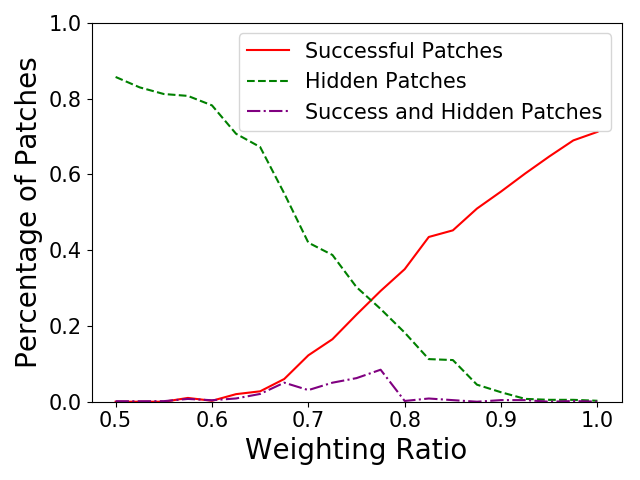}
		\caption{Minimization and attack}
		\label{fig:joint_adaptive_patches}
	\end{subfigure}
	\caption{Heatmap minimization attack. \textbf{(a)} Example of adversarial input minimizing the Grad-CAM output (top); Plot showing the tension between minimization and attack success. \textbf{(b)} Patches minimizing the heatmap and maximizing the attack success respectively (top); Plot of success of the patch vs. minimization of the heatmap.}
\end{figure}

An alternative strategy to fool \tool{} is to minimize the corresponding Grad-CAM region to the greatest extent possible to avoid detection. For example, the adversary can achieve that by modifying the loss function of SGD to optimize for minimal Grad-CAM output. The adversary can start from the adversarial patch image in \Cref{fig:gradcam_minimization} and iteratively add perturbational noise to the region. \Cref{fig:gradcam_minimization}, we show that as our loss converges, the Grad-CAM output is successfully minimized, avoiding detection of the patch.

Our analysis shows that heatmap minimization, i.e., fooling \tool{}, is in tension with the attack success rate of the patch, i.e., fooling $f_m$. \Cref{fig:gradcam_minimization} shows that as more noise is added to the patch, the success rate of the attack is reduced. This result suggests that Grad-CAM is capturing some inherent saliency of the region and cannot be minimized without reducing the attack effectiveness.

The tension between the two objectives could be solved by creating a perturbation that optimizes both for targeted misclassification and for heatmap minimization. Let us consider an adversary that attempts to jointly generate a patch for both criteria by optimizing both objectives during every iteration. Also, to better explore the role played by each objective, we introduce a weighting parameter $\alpha \in [0, 1]$ between the two objectives (i.e., the misclassification objective has weight $\alpha$, and the heatmap minimization has weight $1-\alpha$).
To evaluate then attack, we generate 40 adversarial patches as described by Brown at al.~\cite{brown2017adversarialpatch} for a range of weighting values $\alpha \in [0,1]$. For each patch, we select 100 images from the Imagenet test set~\cite{deng09imagenet} and apply the patch. In total, we generated 4{,}000 malicious images for each weighting ratio $\alpha$. Then, we counted the patches that fool the model, the Grad-CAM\footnote{We use a $<0.5$ overlap ratio between the adversarial pixels and the Grad-CAM region.}, or both.

The results of our experiment are shown in \Cref{fig:joint_adaptive_patches}. As patches work with slight variance, we averaged the attack result to give a better sense of how well a patch would perform. As can be seen in \Cref{fig:joint_adaptive_patches}, there is an inverse relationship between the effectiveness of a patch in fooling $f_m$ (in red) and the region proposal of \tool{} (dashed green), corroborating our previous findings. Additionally, we further plot the patches that successfully fulfill both criteria (in dashed purple), and find that the optimum value is where the ratios of successful misclassifications and hidden patches overlap, giving us patches that fulfill both criteria with an average success rate of 10\%. This strongly suggests that minimizing the Grad-CAM output directly weakens the attack effectiveness and that it is difficult to optimize for both objectives jointly.

\subsubsection{Results on Attacking Region Proposals}

To summarize, in this section, we presented three attacks against the Grad-CAM algorithm. Our evaluation shows that, while the Grad-CAM output can be perturbed with localized perturbations, misredirecting the Grad-CAM heatmap seems an infeasible attack strategy. Also, our results show that minimizing the heatmap size dramatically reduces the effectiveness of the attack. Accordingly, we conclude that attacking the Grad-CAM algorithm may bring a very little advantage to an adversary.

\subsection{Attacking the Class Proposal}
\label{sec:attacking_class_proposal}

Another approach to avoid the localization of the adversarial region is attacking the class proposal step. In this section, we discuss attacks exploiting the algorithms implemented in the class proposal step. In particular, we concentrate on two types of attacks where the adversary intends avoiding detection by generating masks without the adversarial region or weakening the detection by creating inefficient masks.


\subsubsection{Mask Reduction with Specialized Patch Sub-regions} Instead of avoiding detection, the adversary may try to weaken the detection of the patch by reducing the size of the mask of the primary class. The goal of the class proposal is to identify additional classes in the input image to remove ambiguous areas (see \Cref{fig:precise_mask}a-b). Here, the attacker may try tricking the class proposal into identifying classes in sub-regions of the patch. Such an attack results in a mask containing only a portion of the entire patch, thus weakening the response of the model at test time and the detection of the attack.

The creation of a localized universal patch exposing such behavior may be very challenging. Such a patch needs to have a region causing the activation of a class, say $y'$, and, at the same time, another region for the target class $y$. Also, to carry this attack, the patch should cause a change in the prediction of the model when the region for $y'$ is missing. Building a localized universal patch with such a non-linear dependency between the two regions may be hard, if not unfeasible, and we hope other researchers will explore whether such a type of attack is possible.

\subsubsection{Avoiding Mask Reduction by Over-Segmentation} Mask reduction removes from the primary mask areas that are ambiguous, resulting in more efficient masks. Instead of generating inefficient mask by removing malicious regions, the adversary may try to bypass the mask reduction and generate large, ineffective masks like the one shown in \Cref{fig:precise_mask}a. To achieve such a goal, the adversary might attack the image segmentation algorithm, i.e., selective search~\cite{uijlings2013selectivesearch}, and trick the algorithm into returning many small segments outside of the malicious region. When processing small segments, the network may return predictions with low confidence which are ignored when generating the mask for the primary class (see \TopConf{$C$, $k$} in \Cref{alg:class_proposal}). As a result, the final mask will not be refined by removing the ambiguous areas thus weakening the response of the model at test time.

However, attacking the selective search algorithm may not be a feasible strategy. Our inspection of the selective search algorithm revealed that an adversary controlling a limited region of the input image may not be able to influence the generation of a segment of arbitrary size and position. As a result, we believe that attacking the selective search algorithm may be unlikely.

\subsubsection{Results on Attacking the Class Proposals}

To summarize, in this section, we presented three attacks against the algorithms that propose a list of classes present in the image $x$. Our evaluation shows that the goal of avoiding the prediction of the primary class conflicts the goal to hijack the final prediction. Also, our analysis suggests that creating inefficient masks may be either remarkably challenging---if not infeasible---or even not possible. Accordingly, we conclude that attacking the class proposal step may bring little to no advantages to an adaptive adversary.

\subsection{Attacking the Attack Classification}
\label{sec:attacking_classification}

We now consider attacks aimed at fooling the decision procedure of \tool{}.

We now consider attacks aimed at fooling the decision procedure of SentiNet, where extracted regions or inert patterns are overlaid onto benign test images to measure their generalizability. One possible attack in the context of poisoned or trojaned networks has the adversary manipulate the model so that it recognizes the inert pattern as a member of a targeted class. This would let the attacker bypass our defense as adversarial patterns will remain adversarial when overlaid with the``inert'' pattern. This attack does not apply to the setting where an adversary aims to evade an uncompromised network with adversarial patches. For the general case, we note that the choice of inert pattern can be made after the network has been trained, so one can test whether the chosen pattern affects the model. We also found that using patterns made up of various random patterns or noise (see~\Cref{fig:secret_patterns}) produce good results, which further hinders a training-time adversary’s ability to target the specific pattern than will later be used at test-time.
For example, in \Cref{tab:known_attacks_results}, we can see that for the evaluation with the trojaned network, the TP and TN rates of the random noise pattern and checker pattern are within $\leq 0.25\%$.

\section{Discussion}
\label{sec:discussion}

Our results show that our defense is able to record high accuracy metrics for detecting adversarial images in all three cases and is robust to strong adaptive adversaries.  We now discuss in more depth the strengths and limitations of our approach, highlighting some unusual aspects of our design.


\paragraph{Attack Detection} We showed the effectiveness of \tool{} when protecting both compromised and uncompromised networks, without prior knowledge of the network vulnerability. Attack detection rates in four attacks are between 98.5\% and 99.2\%. In one attack, \tool{} scored 85.5\% TP rate, that is comparable with proposed detection techniques~\cite{carlini2018eval}.

\paragraph{Turning a Weakness into a Strength and Proportional Defense} One strength of \tool{} is that it relies on the fact models are compromised by an attack.  Therefore, our detection framework is largely unaffected by the mechanism or deployment the attack uses, detecting attacks successfully as long as they fool the model.  In fact, \tool{} is better at detecting attacks when the adversary is stronger. The properties that characterize a strong attack make it easier for \tool{} detection as such an attack would easily result in outlier behavior outside the threshold of an approximated curve. In real-world conditions, attacks have to be even more robust as they need to tolerate different lighting and viewpoint variations.  Real-world deployments of neural networks could potentially represent the most potent deployment scenario for \tool{}.

\paragraph{Detection of Unsuccessful Attacks} \tool{} can also be further extended with additional functionality.  With the adversarial input detection and class proposal, \tool{} can also analyze the second or subsequent proposed classes, raising the possibility of using \tool{} to detect unsuccessful, attempted attacks. Such a behavior can be useful and help deter attackers from probing and testing our defense with experimental attacks.

\paragraph{Runtime Adversarial Object Suppression} Furthermore, the masking functionality can be used to still preserve functionality by reporting the output label with the inert pattern.  Unlike other detection frameworks which simply identify when an attack has taken place, the calculated mask of \tool{} can easily be used to remove the localized universal attack and salvage as much of the rest of the data as possible. 




\begin{figure}[t]
	\centering
	\includegraphics[width=1\columnwidth]{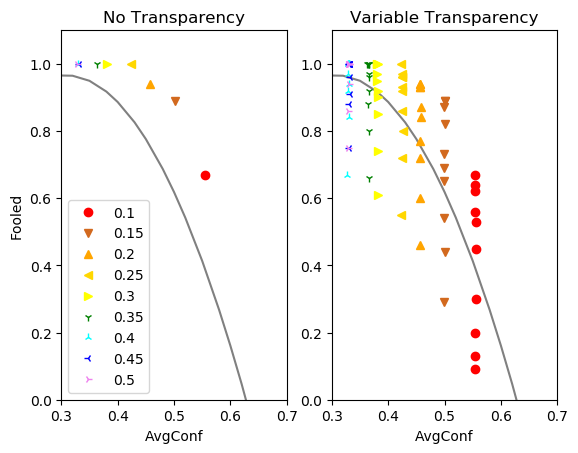}
	\caption{Size analysis of adversarial patches. The legend denotes the ratio of the image to the input, and we can see that the $AvgConf$ drops as the size increases.  On the right, we plot the attack success rate after increasing the transparency values, and we can see that with large ($> 0.35$) patches the input drops below the decision boundary while still retaining a 90\% attack success rate.}
	\label{fig:size_analysis}
\end{figure}

\paragraph{Attack Size} The goal of \tool{} is to capture unreasonably salient attacks, designed to be small and unnoticeable. With this measure, \tool{} largely succeeds at detecting abnormal regions of images that deviate from patches of natural images. When adversaries can modify the majority of pixels in an image, they can override the prediction. As noted by Brown et al.~\cite{brown2017adversarialpatch}, large images of an actual toaster will hijack the same prediction of an adversarial patch at a large enough size, raising an interesting question about what actually constitutes an ``attack''. Based on this observation, we note that the attack's $avgConf$ drops as the size of the object increases, and by increasing the transparency of the object, the attack drops below the threshold while retaining very high attack success (See \Cref{fig:size_analysis}). Similarly as observed by Brown et al.~\cite{brown2017adversarialpatch}, drop of $avgConf$ and high attack success is the result of obfuscating the test inputs with the large targeted class. We conclude that, while our defense correctly captures small patches, the detection of unreasonably large objects is very challenging for \tool{}.

\paragraph{Latency} The runtime of \tool{} is about 2.5s, which is acceptable for tasks like runtime facial recognition. However, it could be impractical for real-time scenarios including autonomous driving. The vast majority of the runtime of \tool{} (1.9s, i.e., ~75\% of the total) is taken by the selective search algorithm. Over the past years, the research community has proposed faster image segmentation approaches; however, none of the new ideas seem to offer the same trade-off between precision, performance, and security as provided by the selective search algorithm. For example, Ren et al.~\cite{ren2015fasterrcnn} and Redmon et al.~\cite{redmon2015yolo, redmon2016yolo9000, redmon2018yolov3} proposes using a neural network to segment images. These new approaches are remarkably fast, taking as little as 0.02 seconds per image (i.e., YOLOv3~\cite{redmon2018yolov3}) to segment an image. While such a low runtime is very desirable for \tool{}, neural network segmentation algorithms introduce severe vulnerabilities in \tool{}. For example, in a recent work, Tram\`{e}r et al.~\cite{tramer2018ad} showed an interesting property of neural network image segmentation techniques, where an adversary can use perturbations to fully control the position and the size of the output segments. Alternatively, in scenarios where efficiency is a priority, a sliding-window approach can be a valid alternative. However, the sliding-window approach has a few drawbacks that make it inadequate for \tool{}. Specifically, it improves execution time only when using large windows, and large windows are more likely to contain multiple classes, affecting the capability to detect adversarial inputs of \tool{}. Overall, based on the current state of the art, we believe selective search is adequate to cope with relevant deployment scenarios of DL systems. Also, we hope that the encouraging results of \tool{} will inspire researchers to propose more efficient and secure segmentation techniques.

\paragraph{Other Classes of Attacks and Future Improvement} The goal of \tool{} is to protect networks against localized universal adversarial inputs, and our results show the viability and robustness of our approach. However, other types of attacks may be possible such as non-universal malicious objects, e.g., patches that work with specific images, or non-localized malicious objects, e.g., disjoint patches. These attacks may pose a challenge for \tool{} and their impact needs to be further studied. 

A promising direction that could improve \tool{} to handle disjoint malicious objects is using other finer-grained visualization techniques, e.g., Grad-CAM++~\cite{chattopadhyay2017gradcamplus}. In general, deep learning interpretability and visualization is a challenging problem and breakthroughs in this area can also allow us to further reason about whether such attacks will take place. 

Handling non-universal patches may require to revisit the anomaly detection phase, which assumes that successful attacks hijack test inputs of different classes. The anomaly detection approach of \tool{} could be extended to take advantage of the richer data provided by the neuron outputs during the inference process. Recent works have shown techniques to detect anomalous high dimensional data using one class neural networks (see, e.g.,~\cite{goldstein2016anomaly,chalapathy2018anomalyoneclass}) which could enhance our current framework.

\section{Related Work}
\label{sec:related_work}

We now review works closely related to this paper. First, we explore the domain of attacks against neural networks. Then, we expand on prior works on localized universal attacks. Finally, we review proposed approaches to detect attacks against neural networks.

\paragraph{Neural Network Attacks} The literature on attacks on neural networks is vast and still growing.  
Szegedy et al.~\cite{szegedy2013boxlbfgs} first demonstrated how adversarial noise can be used to fool neural network classifiers by adding small gradient perturbations to an image that is imperceptible to humans.
Numerous works have built on this approach; some notable works include the Fast Gradient Sign Method~\cite{goodfellow2014fgsm}, DeepFool~\cite{moosavidezfooli2015deepfool}, and Universal Adversarial Perturbations~\cite{moosavidezfooli2016universal}.  
This area of research can be categorized as a cat-and-mouse game in recent years, where defenses are created for new attacks that bypass previous defenses~\cite{carlini2018eval}~\cite{carlini2017tendetection}. Additionally,  attacks certainly are not limited to gradient-perturbation based techniques; data poisoning can be used to cause misintended model behaviors~\cite{shafahi2018poisonfrog}, and compromised hardware can also be used to insert trojans during the network inference procedure~\cite{clements2018hardwaretrojans}.
Akhtar et al.~\cite{akhtar2018adversarialsurvey} provides a useful survey about the current state of the adversarial deep learning field.

\paragraph{Localized Universal Attacks} Multiple works have demonstrated attacks within a variety of classification settings and attacker capabilities. Adversarial patches~\cite{brown2017adversarialpatch} are possibly the most well-known localized universal attack.  These attacks are generated by performing back-propagation on the target class to calculate gradient noise localized to a region of the image. A related attack---Localized and Visible Adversarial Noise~\cite{karmon2018lavan}---operates under a similar principle with smaller but less robust attacks.  Robust Physical-World Attacks on Deep Learning Models~\cite{evtimov2017robust} demonstrates how adversarial perturbations can be disguised as graffiti stickers to fool traffic sign attacks.  Similarly, Sharif et al.~\cite{sharif2016glasses} use perturbations placed on glasses to fool facial classification models.  Trojaned Neural Networks~\cite{Liu2018TrojaningAO} perform back-propagation on specifically chosen neurons in the network rather than the target class. The generated triggers are used to "trojan" the model by performing slight fine-tuning to guide the trigger outputs towards a specified class, making sure that the triggers only cause misclassifications on trojaned models. BadNets~\cite{gu2017badnets} targets traffic sign detection models by inserting pre-chosen patterns into images with the target label, poisoning the data before the training process.

\paragraph{Adversarial Attack Detection} Detection techniques for attacks as a defensive measure have been proposed by many researchers. Safetynets~\cite{lu2017safetynet} is designed to detect adversarial-noise based attacks and exploits the different activations adversarial perturbations produce to train a SVM classifier. Metzen et al.~\cite{metzen2017detectingperturb} use a similar approach by training a modified target classification network to detect adversarial perturbations. Feinman et al.~\cite{feinman2017artifacts} also trains a classifier to detect adversarial perturbational inputs based on the neural network features, while Gong et al.~\cite{gong2017nottwins} introduces a classifier trained to detect adversarially-perturbed images. Magnet~\cite{meng2017magnet} trains a classifier on manifolds of normal examples to detect adversarial perturbations without prior knowledge of the attack. There are also some works designed at creating defenses that do not require training. Grosse et al.~\cite{grosse2017statisticaldetection} uses statistical techniques to distinguish adversarial-perturbations outputs, while Hendrycks et al.~\cite{hendrycks2016earlydetect} uses PCA to visualize differences in perturbed images. A survey by Yuan et al.~\cite{yuan2017attacksanddefenses} covers further detection defenses. All these works are only aimed at defending against adversarial perturbations whereas \tool{} can defend a network against other types of attacks, i.e., data poisoning and trojaning attacks. 

\section{Conclusion}
\label{sec:conclusion}

In this work, we introduce \tool{}, a framework for detecting localized universal attacks on Convolutional Neural Networks.  Our method is notable because it only relies on the malicious behavior of an adversarial attack to perform classifications, without requiring prior knowledge of the attack vector.  We demonstrate the versatility of \tool{} on three experiments with fundamentally different attack mechanisms: a data poisoning attack, a network trojaning attack, and adversarial patches---including physically realizable ones.  We further evaluate the robustness of \tool{} against strong adaptive adversaries by individually testing each component of our defense.  Our approach can be run in realtime in many scenarios, and is flexible and easy to deploy.


We hope \tool{} inspires further approaches towards creating attack-agnostic defenses.  Tailoring a defense towards a specific attack means unknown attacks cannot be captured, and also makes the system highly vulnerable to strong adaptive adversaries. We believe a similar approach can be used to detect other attacks by leveraging the same core concepts of identifying an attack from a model's weakness.


{
    \bibliographystyle{IEEEtranS}
    \bibliography{references}
}

\end{document}